\documentclass[letterpaper,twocolumn,prl,aps,superscriptaddress,amsmath,amssymb,floatfix]{revtex4-1}
\usepackage{mathptmx}
\DeclareMathAlphabet{\mathcal}{OMS}{cmsy}{m}{n}

\usepackage[latin9]{inputenc}
\usepackage{color}
\usepackage{amsmath}
\usepackage{amssymb}
\usepackage{graphicx}
\usepackage{esint}

\usepackage{array}

\usepackage[unicode=true,
 bookmarks=true,bookmarksnumbered=false,bookmarksopen=false,
 breaklinks=false,pdfborder={0 0 1},backref=false,colorlinks=true]
 {hyperref}
\hypersetup{
 linkcolor=magenta,urlcolor=blue,citecolor=blue,pdfstartview={FitH},hyperfootnotes=false}

\makeatletter

\usepackage{textcomp}
\usepackage{epstopdf}

\pdfpageheight\paperheight
\pdfpagewidth\paperwidth

\@ifundefined{textcolor}{}{%
 \definecolor{BLACK}{gray}{0}
 \definecolor{WHITE}{gray}{1}
 \definecolor{RED}{rgb}{1,0,0}
 \definecolor{GREEN}{rgb}{0,1,0}
 \definecolor{BLUE}{rgb}{0,0,1}
 \definecolor{CYAN}{cmyk}{1,0,0,0}
 \definecolor{MAGENTA}{cmyk}{0,1,0,0}
 \definecolor{YELLOW}{cmyk}{0,0,1,0}
}

\usepackage{xcolor}\usepackage{soul}
\newcommand{\ket}[1]{\ensuremath{\left|#1\right\rangle}}

\definecolor{blue}{rgb}{0,0,1}
\definecolor{red}{rgb}{0,0,0}
\definecolor{green}{rgb}{0,0,0}

\newcommand{\red}[1]{\textcolor{red}{ #1}}

\newcommand{\green}[1]{\textcolor{green}{ #1}}

\usepackage{soul}
\newcommand{\Zcr}[1]{{\color{red}#1}}

\makeatother

\begin{document}

\title{Protecting quantum entanglement between error-corrected logical qubits}

% Place the author information here.  Please hand-code the contact
% information and notecalls; do *not* use \footnote commands.  Let the
% author contact information appear immediately below the author names
% as shown.  We would also prefer that you don't change the type-size
% settings shown here.

\affiliation{Center for Quantum Information, Institute for Interdisciplinary Information
Sciences, Tsinghua University, Beijing 100084, China}
\affiliation{Beijing Academy of Quantum Information Sciences, Beijing 100084, China}
\affiliation{CAS Key Laboratory of Quantum Information, University of Science and Technology of China, Hefei 230026, China}
\affiliation{Hefei National Laboratory, Hefei 230088, China}

\author{Weizhou~Cai}
\thanks{These three authors contributed equally to this work.}
\affiliation{Center for Quantum Information, Institute for Interdisciplinary Information
Sciences, Tsinghua University, Beijing 100084, China}
\affiliation{Beijing Academy of Quantum Information Sciences, Beijing 100084, China}

\author{Xianghao~Mu}
\thanks{These three authors contributed equally to this work.}
\affiliation{Center for Quantum Information, Institute for Interdisciplinary Information
Sciences, Tsinghua University, Beijing 100084, China}

\author{Weiting~Wang}
\thanks{These three authors contributed equally to this work.}
\email{wangwt2020@mail.tsinghua.edu.cn}
\affiliation{Center for Quantum Information, Institute for Interdisciplinary Information
Sciences, Tsinghua University, Beijing 100084, China}

\author{Jie~Zhou}
\affiliation{Center for Quantum Information, Institute for Interdisciplinary Information
Sciences, Tsinghua University, Beijing 100084, China}

\author{Yuwei~Ma}
\affiliation{Center for Quantum Information, Institute for Interdisciplinary Information
Sciences, Tsinghua University, Beijing 100084, China}

\author{Xiaoxuan~Pan}
\affiliation{Center for Quantum Information, Institute for Interdisciplinary Information
Sciences, Tsinghua University, Beijing 100084, China}

\author{Ziyue~Hua}
\affiliation{Center for Quantum Information, Institute for Interdisciplinary Information
Sciences, Tsinghua University, Beijing 100084, China}

\author{Xinyu~Liu}
\affiliation{Center for Quantum Information, Institute for Interdisciplinary Information
Sciences, Tsinghua University, Beijing 100084, China}

\author{Guangming~Xue}
\affiliation{Beijing Academy of Quantum Information Sciences, Beijing 100084, China}
\affiliation{Hefei National Laboratory, Hefei 230088, China}

\author{Haifeng~Yu}
\affiliation{Beijing Academy of Quantum Information Sciences, Beijing 100084, China}
\affiliation{Hefei National Laboratory, Hefei 230088, China}

\author{Haiyan~Wang}
\affiliation{Center for Quantum Information, Institute for Interdisciplinary Information
Sciences, Tsinghua University, Beijing 100084, China}

\author{Yipu~Song}
\affiliation{Center for Quantum Information, Institute for Interdisciplinary Information
Sciences, Tsinghua University, Beijing 100084, China}
\affiliation{Hefei National Laboratory, Hefei 230088, China}

\author{Chang-Ling~Zou}
\email{clzou321@ustc.edu.cn}
\affiliation{CAS Key Laboratory of Quantum Information, University of Science and Technology of China, Hefei 230026, China}
\affiliation{Hefei National Laboratory, Hefei 230088, China}

\author{Luyan~Sun}
\email{luyansun@tsinghua.edu.cn}
\affiliation{Center for Quantum Information, Institute for Interdisciplinary Information
Sciences, Tsinghua University, Beijing 100084, China}
\affiliation{Hefei National Laboratory, Hefei 230088, China}

% \date{\today}

\begin{abstract}

{Entanglement represents one of the most important conceptual advances in physics during the last century and is also one of the most essential resources in quantum information science. However, entanglement is fragile and its potential advantages in applications are hindered by decoherence in practice. Here, we experimentally realize entangled logical qubits (ELQ) with a bosonic quantum module by encoding quantum information into spatially separated microwave modes. The entanglement is protected by repetitive quantum error correction, \green{and the coherence time of the purified ELQ via error detection is improved by 45$\%$ compared with the unprotected ELQ and exceeds that of the entangled physical qubits}. \green{In addition, violation of the Bell inequality by logical qubits is demonstrated for the first time with the measured Bell signal $\mathcal{B}=2.250\pm0.019$ after purification, surpassing the classical bound by 13 standard deviations.} The protected ELQ could be applied in future explorations of quantum foundations and applications of quantum networks. }

\end{abstract}

\maketitle

%\section{Introduction}
%\label{sec:introduction}

In 1935, Einstein, Podolsky, and Rosen (EPR) conceived a gedanken experiment that challenged the completeness of quantum mechanics~\cite{Einstein1935}. The curiosity about the description of Nature has stimulated further controversy, and the celebrated Bell inequality was proposed for testing the foundation of quantum mechanics~\cite{Bell2004}. Since then, great experimental efforts have been dedicated to excluding the hidden-variable theory\textcolor{red}{~\cite{Clauser1969,Aspect1981,Zeilinger1998,Hill2022}}, and the concept of entanglement has been validated for nonlocal realism interpretation. Currently, entanglement is found to be indispensable when describing the world at the microscopic scale~\cite{Vedral2014}, such as the photosynthetic dynamics of biologic systems~\cite{Lambert2013} and exotic phases of quantum matter~\cite{Laflorencie2016}, and might also explain the classical-to-quantum transition of macroscopic objects~\cite{Preskill2022}. More recently, the conjecture ``ER (Einstein-Rosen Bridge)=EPR" was proposed~\cite{Maldacena2013}, and entanglement is the key to connecting quantum theory and gravity.

\begin{figure*}
\includegraphics[scale=1]{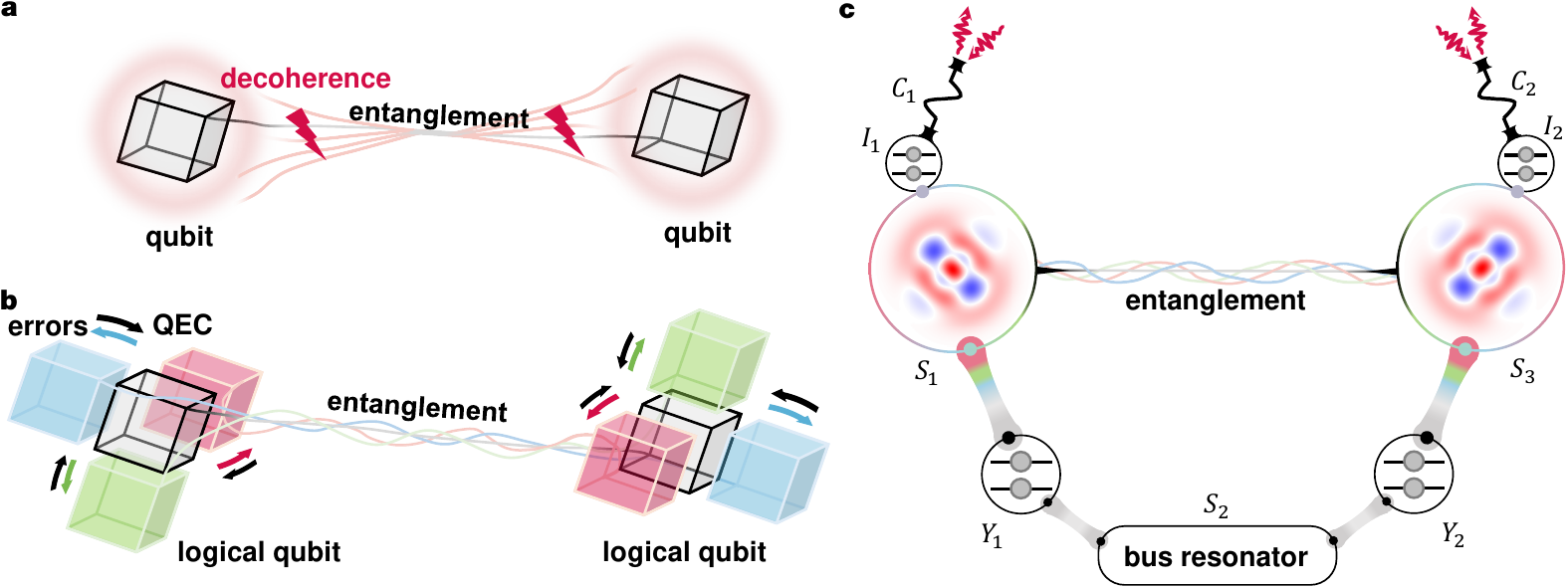}
\caption{\textbf{Principle and setup for entangled logical qubits (ELQ).}  \textbf{a} and \textbf{b} The effect of decoherence on quantum entanglement. For physical qubits \textbf{a}, the environment (red shadows) could degrade the entanglement between the two qubits (white cubes). For logical qubits \textbf{b} in the code subspaces of the quantum error correction (QEC) codes, local noise would project the entanglement between code-code subspaces into code-error or error-error subspaces (colored cubes). The entanglement between the original \red{code-code} subspaces can be recovered by QECs. \textbf{c} Schematic of a superconducting system for a bosonic quantum module of logical qubits. Quantum information is encoded into bosonic modes ($S_1$ and $S_3$), whose entanglement is generated via two auxiliary transmon qubits ($Y_1$ and $Y_2$) and a bus resonator ($S_2$). The QEC operations and readouts of the ELQ are implemented by the control transmon qubits $I_1$ and $I_2$.}
\label{Fig1}
\end{figure*}

The exploration of entanglement not only contributes to the foundation of quantum mechanics but also drives the development of quantum information technology~\cite{Nielsen,Horodecki2009}. Entanglement allows many counterintuitive protocols, such as quantum teleportation for transferring unknown quantum states~\cite{Bennett1993}, entanglement swapping for entangling noninteracting particles~\cite{ifmmodeZelseZfiukowski1993}, and precision measurement beyond the shot-noise limitation~\cite{Degen2017,Pirandola2018}. Therefore, entanglement becomes \red{the most important quantum resource} for applications, including quantum computation, communication, and sensing. For instance, the exponential speed-up of a universal quantum computer compared to a classical computer comes from the highly entangled qubits~\cite{Nielsen}. %The quantum error correction technique requires entanglement to share the quantum information among many qubits to fight against local noise. 

However, entanglement is highly fragile. Although entanglement cannot be generated by local  operations and classical communications~\cite{Nielsen}, it can be destroyed by local decoherence effects. Figure~\ref{Fig1}a illustrates a generic setup of entanglement, where two localized qubits share nonlocal quantum features. However, the uncontrollable environment (red shadow) interacting with each qubit would destroy the entanglement due to the leakage of quantum information to the environment. Thus, decoherence imposes an obstacle to entanglement-assisted or entanglement-enhanced quantum information processing in practice. A critical challenge is to protect the entanglement.

Here, we realize the protection of quantum entanglement between logical qubits by the quantum error correction (QEC) technique. Compared to a physical qubit that supports only two-dimensional Hilbert space, \red{a logical qubit is in a two-dimensional code subspace of a high-dimensional quantum system, and the redundant degrees of freedom allow the detection and correction of potential errors that project the logical qubit into the error subspaces.} Therefore, the quantum information stored by logical qubits is more robust to local noise. As shown in Fig.~\ref{Fig1}b, by replacing each physical qubit in Fig.~\ref{Fig1}a with a logical qubit, the system-environment interactions convert entanglement between code-code subspaces to code-error or error-error subspaces instead of directly destroying it, and entanglement between logical qubits can be recovered by local QEC operations.

% \section{Results}

% \subsection{Principle}

In a hardware-efficient circuit quantum electrodynamics platform~\cite{Blais2020cQED,BlaisNP2020}, ELQ based on a bosonic quantum module is experimentally investigated, as schematically shown in  Fig.~\ref{Fig1}c. The fundamental modes of two 3D coaxial cavities~\cite{Paik2011,ReagorAPL2013} ($S_1$ and $S_3$) are used for storing quantum information. These two bosonic modes are spatially separated with no direct interaction between them and their entanglement can be generated by two auxiliary transmon qubits~\cite{Koch2007} ($Y_1$ and $Y_2$) and a bus resonator ($S_2$). Each of the two modes also couples to a control qubit ($I_1$ and $I_2$ respectively), which assists the manipulation and readout of the corresponding mode. %\red{by coupling to an external channel ($C_1$ and $C_2$ respectively)}. 
For more details about the experimental device and setup, see Ref.~\cite{Supplement}. Such a module not only serves as local quantum memory for storing quantum information but also provides the simplest node for quantum networks. Since the electromagnetic wavepacket mode offers an excellent flying carrier to distribute quantum information over a long distance, entanglement among modules can be realized by \red{pitching and catching} the propagating wavepackets in the external communication channels \red{($C_1$ and $C_2$)} ~\cite{Pfaff2017,Axline2017}.

For both the stored and propagating bosonic modes, the dominant decoherence effect is the excitation loss error due to local energy decay or propagation attenuation. For a physical qubit that encodes quantum information to the lowest excitation levels of such a mode, i.e. the Fock states $\{|0\rangle,\,|1\rangle\}$, the error induces a mixed quantum state because the contributions to the population of $|0\rangle$ by either the initial state or the error cannot be distinguished. Therefore, the entanglement between physical qubits suffers from error (Fig.~\ref{Fig1}A). Taking advantage of the high-dimensional Hilbert space of bosonic modes, logical qubits utilizing higher Fock states based on a binomial QEC code~\cite{Michael2016,Hu2019} are adopted to overcome the photon loss error, with the logical basis states being superpositions of the Fock states with even parity:
\begin{equation}
\left|0_{\mathrm{L}}\right\rangle=\frac{|0\rangle+|4\rangle}{\sqrt{2}},\ \left|1_{\mathrm{L}}\right\rangle=|2\rangle. 
\label{eq:Basis_States}
\end{equation}
When a single-photon-loss error occurs, the logical state is projected into the error subspace with the corresponding projected basis states $\left|0_{\mathrm{E}}\right\rangle=|3\rangle$ and $\left|1_{\mathrm{E}}\right\rangle=|1\rangle$ with an odd parity. The error subspace is clearly orthogonal to the code subspace, and thus the error can be detected through a local parity measurement $\mathcal{P}=e^{i\pi a^{\dagger}a}$~\cite{SunNature2014} with $a^{\dagger}$ and $a$ denoting the creation and destroy bosonic operators of the mode. %Consequently, the ambiguity of the error could be eliminated according to the parity $\mathcal{P}=\pm1$, 
Furthermore, the corrupted quantum state in the error space ($\mathcal{P}=-1$) can be recovered to the code subspace ($\mathcal{P}=1$) by an error correction operation $\mathcal{R}$ with $\mathcal{R}\left|0_{\mathrm{E}}/1_{\mathrm{E}}\right\rangle=\left|0_{\mathrm{L}}/1_{\mathrm{L}}\right\rangle$. The error detection and correction operations on the two logical qubits could be separately implemented through local interactions assisted by the two corresponding control qubits ($I_1$ and $I_2$ in Fig.~\ref{Fig1}C). These local operations cannot bring entanglement or correlations into the system, and thus the protection of the entanglement is built on the protection of each single logical qubit by QEC.

%\red{(separately repetitive QEC protects single logical qubit)} 
% \subsection{Single logical qubit}
In our module, the two storage cavity modes can be manipulated individually without perturbing each other, and the performance of each bosonic logical qubit is characterized separately. Figure~\ref{Fig2} shows the results for cavity $S_1$ as an example. Employing the nonlinear interaction between the cavity mode and the coupled transmon qubit, arbitrary operations of the system could be realized by external drives on them. \red{In our experiment, the encoding, decoding, and gates of the binomial code are achieved with numerically optimized drive pulses~\cite{Heeres2017,Hu2019}. Furthermore,} the error syndrome measurement and the corresponding recovery operation are simultaneously implemented by a single pulse to avoid the electronic latency of the real-time feedback control~\cite{Ma2020NP,Wang2022NC}. This autonomous QEC transfers the error entropy on the logical qubit to the control qubit. \red{For example, if an error occurs the error information is stored in the control qubit (the qubit is flipped to the excited state) while the corrupted logical qubit is restored.} More experimental details for the operations as well as the results for cavity $S_3$ are provided in Ref.~\cite{Supplement}.

\begin{figure}
\centering
\includegraphics[scale=1]{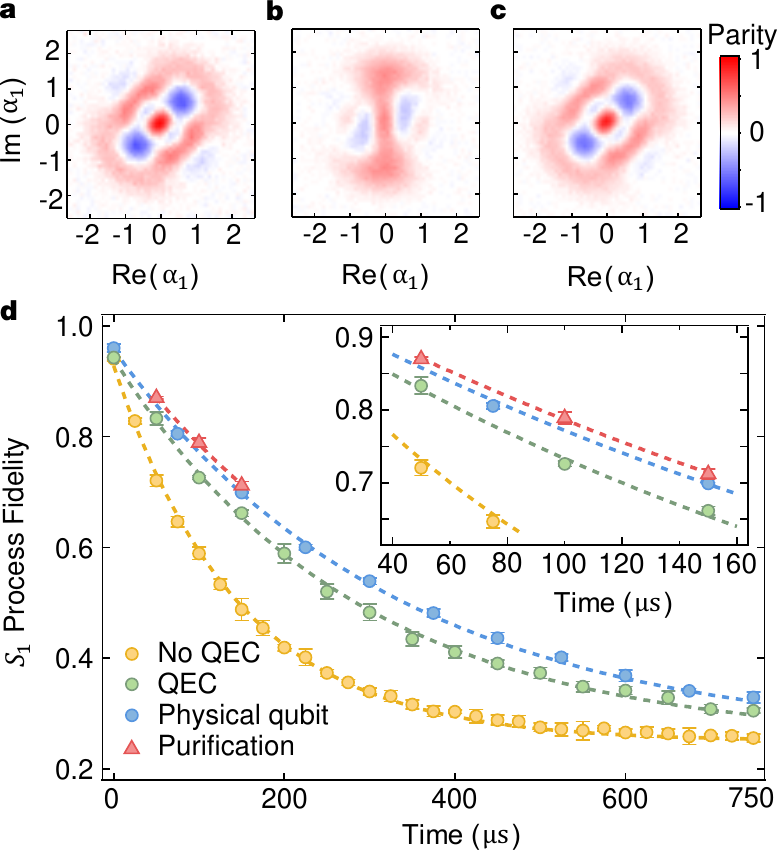}
\caption{\textbf{Performance of a single logical qubit.}  \textbf{a-c} The Wigner tomography plots of the logical qubit ($S_1$) at the initial state $(\ket{0_{\mathrm{L}}}-i\ket{1_{\mathrm{L}}})/\sqrt{2}$ \textbf{a}, after $50\,\mu$s evolution without \textbf{b} and with \textbf{c} QEC protection, respectively.  \textbf{d} Process fidelities of quantum information encoded in cavity $S_1$ as a function of time for different schemes. \red{The orange circles, green circles, and pink triangles} are the results of the binomial code without QEC, with repetitive QEC, and with purification via error detection, respectively. The blue circles show the results of the physical qubit encoded with the lowest Fock states. \red{The error bars correspond to the standard deviation from multiple measurements. The dashed lines are the fit with an exponential decay function.} Inset: A zoomed-in plot of \textbf{d}.%The green dot line represents the binomial code protected by repetitive AQEC processes with 50$\mu$s time interval. The process fidelity of the AQEC-protected binomial codes has a huge increase than the code without the AQEC protection. With the experimental data post-processing, the error spaces are dropped from the data of the deterministic AQEC process (green dot line), which makes the experimental results have better performance and is shown as the purple hollow circle. The blue dot line represents the process fidelity changed with the time of Fock 0 and 1 encoding without any protection. The fuchsin hollow triangle represent the binomial code under the repetitive error syndrome detection without error correction operations in the process, which has a better performance than the Fock 0,1 encoding and is the so-called break-even point. 
}
\label{Fig2}
\end{figure}

The protection of logical qubits \red{with repetitive QEC} is first illustrated for a given initial state$(\ket{0_{\mathrm{L}}}-i\ket{1_{\mathrm{L}}})/\sqrt{2}$, whose Wigner function distribution is shown in Fig.~\ref{Fig2}a. The purity of the state after a duration of $50\,\mu$s degrades due to the photon loss error, as seen by the reduced contrast of the distribution in Fig.~\ref{Fig2}b. However, by implementing a QEC operation, the logical state is recovered, as shown in Fig.~\ref{Fig2}c. The ability to preserve arbitrary quantum states by the logical qubit is quantified by the process fidelity $F$, which decays with the total storage time $t$. The results are summarized in Fig.~\ref{Fig2}d. Compared with the physical qubit encoded in $\{\ket{0},\ket{1}\}$, the logical qubit without QEC protection shows a lower $F$ since the excitation loss error rate is larger for the binomial code with a higher excitation number. By applying repetitive QEC with an interval of $50\,\mu$s, $F$ is significantly improved to approach that of the physical qubit. Fitting the curves with $F(t)=\frac{1}{4}+\frac{3}{4}A\mathrm{e}^{-t/T_1}$, with $A\leq1$ being limited by imperfect encoding and decoding operations, the resulting coherence times are $T_1=147(2),\,280(4)$, and $328(7)\,\mu$s for the logical qubit without QEC, with QEC, and the physical qubit, respectively. These results confirm that QEC indeed protects the logical qubits by improving the coherence time by a factor of $90\%$.
 	 
%\red{(separately repetitive QEC for extending the lifetime of the entanglement)} 
% \subsection*{Entangled logical qubits (ELQ)}
With the ability to implement high-performance repetitive QEC on a single logical qubit, protecting entanglement between two logical qubits via QEC is then investigated. \red{The ELQ is realized by encoding each logical qubit ($S_1$, $S_3$) with its adjacent auxiliary qubit ($Y_1$, $Y_2$), while the two auxiliary qubits are initially prepared in the Bell state $\Psi^{+}=(\ket{0}\ket{1}+\ket{1}\ket{0})/\sqrt{2}$ through the bus cavity $S_2$.} %\blue{The ELQ is realized by swapping the states between the two logical qubits and the two auxiliary qubits ($Y_1$ and $Y_2$), which are prepared to the Bell state $\Psi^{+}=(\ket{0}\ket{1}+\ket{1}\ket{0})/\sqrt{2}$ through the bus cavity $S_2$.} 
Figure~\ref{Fig3}a represents the joint Wigner function distribution of the two logical qubits, and the correlation between the quadratures of the separated cavity modes reveals the entanglement~\cite{Wang2016,Supplement}. After a $50\,\mu$s waiting time, the correlation decays due to decoherence (Fig.~\ref{Fig3}b) \red{but is largely} recovered by separated QEC operations on each logical qubit (Fig.~\ref{Fig3}c). The protection of entanglement is further quantified by the logical Bell state fidelity, as shown in Fig.~\ref{Fig3}d, which also exponentially decays with time. \textcolor{red}{Comparing the situations of ELQ with and without repetitive QEC, the \red{entanglement decay time} is improved by 45\% from $82(4)\,\mu$s to $119(3)\,\mu$s.} 

\begin{figure*}
\includegraphics[scale=1]{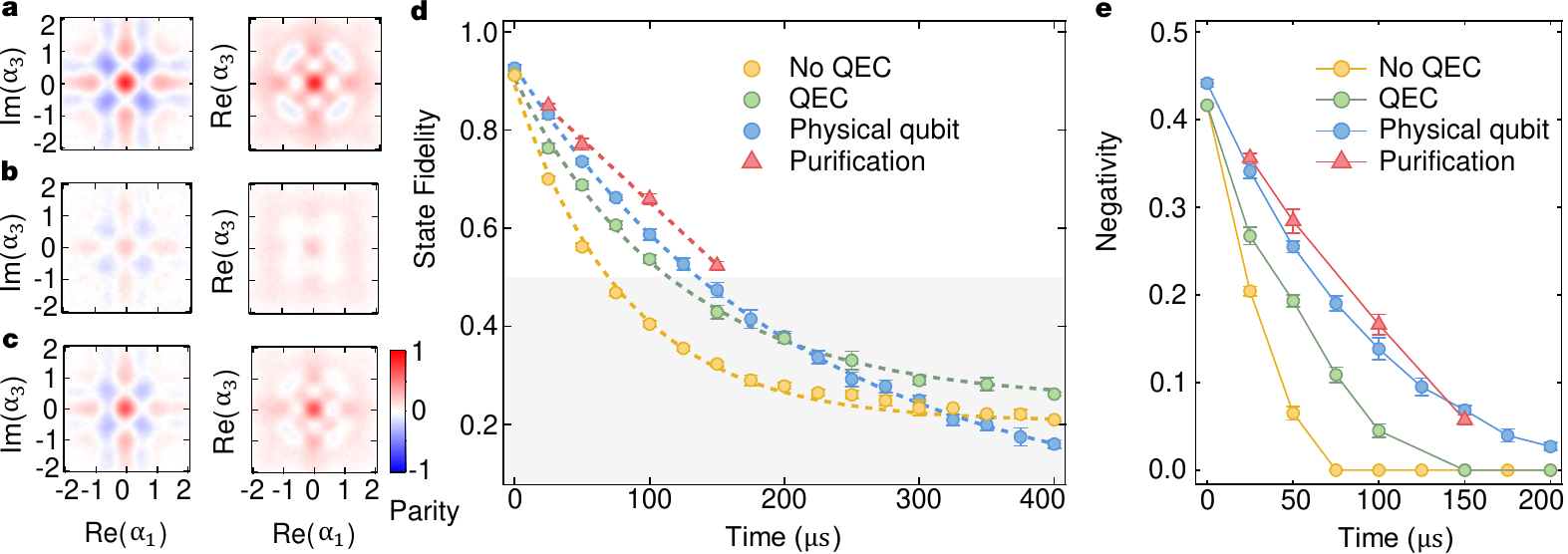}
\caption{\textbf{Performance of the entangled logical qubits.} \textbf{a-c} Wigner tomography plots of the ELQ at the initial state $(\ket{0_L}\ket{1_L}+\ket{1_L}\ket{0_L})\sqrt{2}$ \textbf{a}, after $50\,\mu$s evolution without \textbf{b} and with \textbf{c} QEC protection, respectively. \textbf{d} Fidelityies of Bell state $(\ket{0}\ket{1}+\ket{1}\ket{0})\sqrt{2}$  \red{as a function of time} with different encodings or different quantum error processings of the ELQ. \red{The orange circles, green circles, and pink triangles} are the results of the ELQ without QEC, with \red{repetitive} QEC, and with purification via error detection, respectively. The blue circles show the results of the physical qubits encoded with the lowest Fock states. \red{The dashed lines are the fit with an exponential decay function}.
%two entangled cavity S1 and S3. The yellow dot line shows the entangled binomial codes suffering the photon loss errors without the protection of the QEC process. The green dot line represents the entangled binomial codes protected by repetitive QEC processes with 50$\mu$s time interval, with inserting two more points of QEC processes at 25$\mu$s and 75$\mu$s. The state fidelity of the QEC-protected entangled binomial codes has a prominent increase than the codes without the QEC protection. With the experimental data post-processing, one of the four possible situations, the no photon loss occurs on each code, is picked from the deterministic QEC process (green dot line). It makes the experimental results have better performance and is shown as the purple hollow circle. The blue dot line represents the Bell state fidelity of Fock 0 and 1 encoding changed with time without any protection. The fuchsin hollow triangle represent the entangled binomial code under the repetitive error syndrome detection without error correction operations in the processes, which has a better performance than the Fock 0,1 encoding. %The insets at the top, right, and bottom-left represents the Wigner tomography of the entangled binomial state $((\ket{0}+\ket{4})\ket{2}+\ket{2}(\ket{0}+\ket{4}))/2\sqrt{2}$, after the encoding, at 50$\mu$s with QEC, and at 50$\mu$s without QEC, respectively. 
\textbf{e} Negativity of the ELQ \red{as a function of time} under different situations in \textbf{d}. \red{The error bars in \textbf{d} and \textbf{e} correspond to the standard deviation from multiple measurements.}}
\label{Fig3}
\end{figure*}

\red{The performances of the measured single logical qubits and ELQ are lower than those of the physical qubits.} The main obstacle for current logical qubits to achieve the break-even point is the imperfect recovery operations on the logical qubits, which are repetitively implemented during experimental cycles~\cite{Supplement,Hu2019,Xu2022}. 
%\Zcb{Note, entanglement is resource, so the probabilistic preparation of high-fidelity state is useful}
Considering the practical applications for preparing Bell states, we might improve the fidelity and the performance of the ELQ by a purification strategy that discards the corrupted logical qubits. Since the errors could be detected by simple syndrome measurements, the limitation of imperfect $\mathcal{R}$ could be circumvented by postselecting the state only when no single-photon-loss error occurred. As shown by the red triangles in Fig.~\ref{Fig2}d, the purified logical qubit shows a further improved coherence time by error detections and \Zcr{discarding the corrupted states without recovery operations}, confirming the effectiveness of purification and revealing the limitation due to the operation error. Through purification, the ELQ is also protected with a lifetime even longer than the physical qubits (Fig.~\ref{Fig3}d triangles), demonstrating the advantage of the logical qubits.

%\Zcr{For dislodging these operation errors, we repetitively measure the error syndrome (parity)~\cite{SunNature2014} to continuously track the quantum trajectory of each single experiment and then decode the bosonic code from the cavity to the ancilla qubit at the end of the process. With post-selecting code spaces of the logical qubit in each parity measurement~\cite{Supplement}, we obtain better process fidelity of a purified logical qubit than of an error-corrected logical qubit. The experiment results are summarized in Fig.~\ref{fig:repetitive_QEC}, where the triangles represent the process fidelity of the purified logical qubit.}

As a nonlocal quantum resource, entanglement might exist even when the Bell state fidelity is lower than 0.5 since fidelity degradation might also be attributed to local unitary operation. Therefore, the negativity $\mathcal{N}$~\cite{Vidal2002PRA} as a measure of entanglement is introduced to benchmark the protection of entanglement, and its change over time is shown in Fig.~\ref{Fig3}E. According to the criteria that entangled quantum states have $\mathcal{N}>0$, our result shows that the lifetime of entanglement for the ELQ is improved by over $33\%$ from less than $75\,\mu$s to more than $100\,\mu$s by QEC. In addition, the purified ELQ has performances comparable or better than the physical qubit encoding. Our experiments on ELQ also provide an alternative approach for quantum entanglement purification via QEC~\cite{Dur2007}, which could be useful for next-generation quantum repeaters~\cite{Azuma2022}. %\Zcb{which indicates that with repetitively transferring of error entropy from the system to the environment, the entanglement between two error-correctable logical qubits can be protected by QEC.} [3rd generation quantum repeater]

\begin{figure}
\centering
\includegraphics[scale=1]{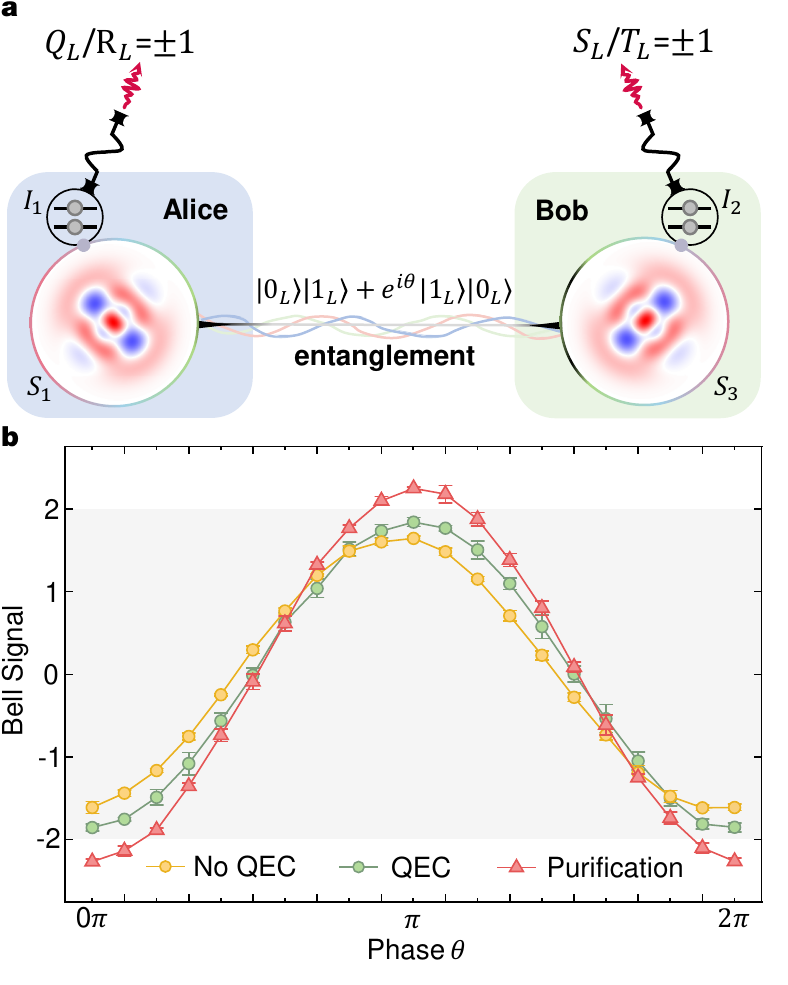}
\caption{\textbf{Bell test of the entangled logical qubits.} \textbf{a} Schematic experimental setup for the Bell inequality test of ELQ. \textbf{b} The Bell test results against the relative phase ($\theta$) of the Bell states at 25$\,\mu$s. The orange and green circles show the results of ELQ with and without QEC protection, respectively. \green{The pink triangles are the results purified via error detection with the maximum $\mathcal{B}=2.250\pm0.019$, surpassing the classical bound (shaded region) and  demonstrating the violation of the Bell inequality.} %The shaded region represents the classically allowable region.
}
\label{Fig4}
\end{figure}

%(\textbf{B}) A zoom-in plot of bell signal, which showing the violation of classical limit by ELQ under error detection when phase $\theta=\pi$. \textbf{(C)} The corresponding negativity of ELQ with different experimental configurations at $25\,\mu $s.

Finally, the ELQ is applied to the test of Bell inequality to verify quantum mechanics (Fig.~\ref{Fig4}a). The Bell operator $\mathcal{B}=Q_LS_L+R_LS_L+R_LT_L-Q_LT_L$ is measured by changing the local measurement bases ${Q_L,R_L}$ and ${S_L,T_L}$ for the two logical qubits [see Ref.~\cite{Supplement} for more details]. The ELQ is prepared to a maximally entangled state $(\ket{0_{\mathrm{L}}}\ket{1_{\mathrm{L}}}+e^{i\theta}\ket{1_{\mathrm{L}}}\ket{0_{\mathrm{L}}})/\sqrt{2}$ with a relative phase $\theta$, and Bell-test experiments are performed for ELQ after a storage time of $25\,\mu$s, \red{which also imitates a potential propagation loss for the ELQ traveling over a long distance in a quantum network}. The experimental results are summarized in Fig.~\ref{Fig4}b, where QEC enhances the Bell signal and a violation of the Bell inequality is realized by the purified ELQ. \green{The maximum of the measured Bell signals $\mathcal{B}=2.250\pm0.019$ surpasses the classical bound by 13 standard deviations.}

In conclusion, entangled logical qubits and the protection of their entanglement are demonstrated in a bosonic quantum module for the first time. As an important resource for quantum information science, the maximally entangled logical state is protected, with its lifetime extended to be longer than that of the entangled physical qubits via purification. \red{Therefore, with the verified advantage of logical qubits, \green{the bosonic quantum module offers a basic building block for scalable quantum information processors and networks~\cite{Bosonicreview,Bosonicreview_Joshi_2021}.}} Our work appeals to further efforts to extend error-corrected logical qubits to multipartite entanglement and to prepare entangled states fault-tolerantly. Taking advantage of the bosonic modes that combat the excitation loss errors of photons, the entanglement generated by our module could be distributed over a long distance for potential applications in quantum networks and distributed quantum sensing. Based on that, a new generation of experimental testing of quantum foundations based on logical qubits could also be anticipated.

% \section{Acknowledgments}

\textbf{Acknowledgments:}
This work was supported by the National Natural Science Foundation of China (Grants No. 92165209, No. 11925404, No. 12204052, No. 11890704, 92265210, and No. 12061131011), Natural Science Foundation of Beijing (Grant No. Z190012), Fundamental Research Funds for the Central Universities, China Postdoctoral Science Foundation (BX2021167), Grant No. 2019GQG1024 from the Institute for Guo Qiang, Tsinghua University, National Key Research and Development Program of China (Grants No. 2017YFA0304303), and Key-Area Research and Development Program of Guangdong Province (Grant No. 2020B0303030001). This work was partially carried out at the USTC Center for Micro and Nanoscale Research and Fabrication. 

\textbf{Author contributions:}
L.S. and C.-L.Z. conceived the experiments. W.C., X.M. and W.W. performed the experiments and analyzed the data with the assistance of J.Z., Y.M., X.P., Z.H., and X.L. L.S. directed the project. C.-L.Z. provided the theoretical support. W.C., X.M., W.W., and J.Z. performed the numerical simulations. W.C., H.W, and Y.S. fabricated the 3D cavity. G.X. and H.Y. fabricated the tantalum transmon qubits and provided further experimental support. W.C., X.M., W.W., J.Z., C.-L.Z., and L.S. wrote the manuscript with feedback from all authors. 

\textbf{Competing}: The authors declare no competing interests. 

\textbf{Data and materials availability}: All data are available in the main text or the supplementary materials.

%Control: production of eprint (-1) disabled
%

%\bibliography{bibliography}
%\bibliographystyle{Zou}

%\end{document}

%\clearpage{}

\cleardoublepage{}

\onecolumngrid 
\global\long\def\thefigure{S\arabic{figure}}%
\setcounter{figure}{0} 
\global\long\def\thepage{S\arabic{page}}%
\setcounter{page}{1} 
\global\long\def\theequation{S.\arabic{equation}}%
\setcounter{equation}{0} %\renewcommand{\thesection}{S.\Roman{section}}
\setcounter{section}{0}
\begin{center}
	\textbf{\Large{}Supplementary Materials for ``Protecting quantum entanglement between error-corrected logical qubits"}{\Large\par}
	\par\end{center}
%%\documentclass [prl,aps,letterpaper,preprint,amsmath,amssymb,floatfix,superscriptaddress] {revtex4}
%\documentclass [prL,aps,letterpaper,twocolumn,superscriptaddress,amsmath,amssymb,floatfix] {revtex4}
%\pdfoutput=1
%\usepackage{graphicx}
%\usepackage{textcomp}
%\usepackage{mathptmx} % makes Times Roman font for text AND math
%\usepackage{amsmath}
%\usepackage{xcolor}
%\usepackage{soul}
%\usepackage{array}
%\usepackage[unicode=true,bookmarks=true,bookmarksnumbered=false,bookmarksopen=false,breaklinks=false,pdfborder={0 0 1},backref=false,colorlinks=true]{hyperref}
%\hypersetup{linkcolor=magenta,urlcolor=blue,citecolor=blue,pdfstartview={FitH},hyperfootnotes=false}
%
%% define macros
%\newcommand{\dg}{$^\circ$ }
%\newcommand{\dgc}{$^\circ\mathrm{C}$}
%\newcommand{\bra}[1]{\ensuremath{\left\langle#1\right|}}
%\newcommand{\ket}[1]{\ensuremath{\left|#1\right\rangle}}
%
%\renewcommand{\thefigure}{S\arabic{figure}}
%\renewcommand{\thetable}{S\arabic{table}}
%
%%\renewcommand\thesection{\Alph{section}}
%%\newcommand{\onlinecite}[1]{\hspace{-1 ex} \nocite{#1}\citenum{#1}}
%
%\definecolor{blue}{rgb}{0,0,1}
%\definecolor{red}{rgb}{0,0,0}
%\definecolor{green}{rgb}{0,1,0}
%\newcommand{\red}[1]{\textcolor{red}{ #1}}
%\newcommand{\blue}[1]{\textcolor{blue}{ #1}}
%\newcommand{\green}[1]{\textcolor{green}{ #1}}

%\begin{document}
	
	\title{Supplementary Materials for ``Protecting quantum entanglement between error-corrected logical qubits"}
	
	\affiliation{Center for Quantum Information, Institute for Interdisciplinary Information
		Sciences, Tsinghua University, Beijing 100084, China}
	\affiliation{Beijing Academy of Quantum Information Sciences, Beijing 100084, China}
	\affiliation{CAS Key Laboratory of Quantum Information, University of Science and Technology of China, Hefei 230026, China}
	\affiliation{Hefei National Laboratory, Hefei 230088, China}
	
	\author{Weizhou~Cai}
	\thanks{These three authors contributed equally to this work.}
	\affiliation{Center for Quantum Information, Institute for Interdisciplinary Information
		Sciences, Tsinghua University, Beijing 100084, China}
	\affiliation{Beijing Academy of Quantum Information Sciences, Beijing 100084, China}
	
	\author{Xianghao~Mu}
	\thanks{These three authors contributed equally to this work.}
	\affiliation{Center for Quantum Information, Institute for Interdisciplinary Information
		Sciences, Tsinghua University, Beijing 100084, China}
	
	\author{Weiting~Wang}
	\thanks{These three authors contributed equally to this work.}
	\email{wangwt2020@mail.tsinghua.edu.cn}
	\affiliation{Center for Quantum Information, Institute for Interdisciplinary Information
		Sciences, Tsinghua University, Beijing 100084, China}
	
	\author{Jie~Zhou}
	\affiliation{Center for Quantum Information, Institute for Interdisciplinary Information
		Sciences, Tsinghua University, Beijing 100084, China}
	
	\author{Yuwei~Ma}
	\affiliation{Center for Quantum Information, Institute for Interdisciplinary Information
		Sciences, Tsinghua University, Beijing 100084, China}
	
	\author{Xiaoxuan~Pan}
	\affiliation{Center for Quantum Information, Institute for Interdisciplinary Information
		Sciences, Tsinghua University, Beijing 100084, China}
	
	\author{Ziyue~Hua}
	\affiliation{Center for Quantum Information, Institute for Interdisciplinary Information
		Sciences, Tsinghua University, Beijing 100084, China}
	
	\author{Xinyu~Liu}
	\affiliation{Center for Quantum Information, Institute for Interdisciplinary Information
		Sciences, Tsinghua University, Beijing 100084, China}
	
	\author{Guangming~Xue}
	\affiliation{Beijing Academy of Quantum Information Sciences, Beijing 100084, China}
	\affiliation{Hefei National Laboratory, Hefei 230088, China}
	
	\author{Haifeng~Yu}
	\affiliation{Beijing Academy of Quantum Information Sciences, Beijing 100084, China}
	\affiliation{Hefei National Laboratory, Hefei 230088, China}
	
	\author{Haiyan~Wang}
	\affiliation{Center for Quantum Information, Institute for Interdisciplinary Information
		Sciences, Tsinghua University, Beijing 100084, China}
	
	\author{Yipu~Song}
	\affiliation{Center for Quantum Information, Institute for Interdisciplinary Information
		Sciences, Tsinghua University, Beijing 100084, China}
	\affiliation{Hefei National Laboratory, Hefei 230088, China}
	
	\author{Chang-Ling~Zou}
	\email{clzou321@ustc.edu.cn}
	\affiliation{CAS Key Laboratory of Quantum Information, University of Science and Technology of China, Hefei 230026, China}
	\affiliation{Hefei National Laboratory, Hefei 230088, China}
	
	\author{Luyan~Sun}
	\email{luyansun@tsinghua.edu.cn}
	\affiliation{Center for Quantum Information, Institute for Interdisciplinary Information
		Sciences, Tsinghua University, Beijing 100084, China}
	\affiliation{Hefei National Laboratory, Hefei 230088, China}

	%\date{\today}
	%\pacs{} 
	
	\maketitle
	
	\tableofcontents

	\section{Experimental device and setup}
	
	\subsection{Device structure}
	
\begin{figure}
		\centering
		\includegraphics[scale=1]{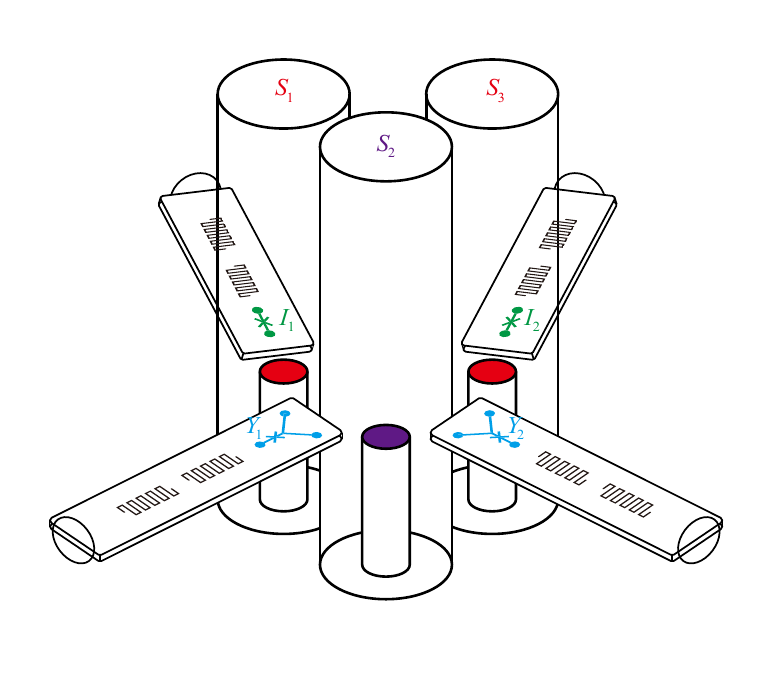}
		\caption{\textbf{Schematic of the real device.} The main part of the device is made from 5N5 aluminum with a purity of 99.9995\%. This device holds three ${\lambda}/{4}$ coaxial cavities ($S_1$, $S_2$, and $S_3$), two of which ($S_1$ and $S_3$) are used as the logical qubits. Four trenches are used to host the qubit chips. Both chip designs ($Y$- and $I$-shaped) have an additional stripline resonator acting as a bandpass Purcell filter.}
		\label{fig:SM_SolidWorks}
	\end{figure}

	Our experimental device consists of three 3D-coaxial cavities ($S_1, S_2$, and $S_3$)~\cite{Reagor2016PRB}, two $Y$-shaped transmon qubits ($Y_1$ and $Y_2$), and two $I$-shaped transmon qubits ($I_1$ and $I_2$), as shown in Fig.~\ref{fig:SM_SolidWorks}. The whole package is made from 5N5 high-purity aluminum (99.9995\%), inside which the three coaxial cavities are machined together with four trenches for the qubit chips. Each transmon qubit couples to its individual Purcell filter and readout resonator, and all of them  are fabricated on a single chip. When combined with the corresponding output cable, the readout resonator can also be extended as a communication channel for transferring information in a quantum network, as indicated in Fig.~1(c) of the main text. Each of the three coaxial cavities works as a $\lambda/4$ transmission line resonator, with one end being grounded and the other end being open. The outer diameter of all three coaxial cavities is 9.6~mm and the inner diameter is 3.2~mm.
	The resonance frequency is determined by the length of the inner conductor: the lengths for the two logical qubits $S_1$ and $S_3$ are 10.6~mm and 10.8~mm respectively and the length for the other common cavity $S_2$ is 11.8~mm.
	In the upper part of the coaxial cavity, the electromagnetic mode is exponentially suppressed into the circular waveguide.
	The length of the circular waveguide is designed to be long enough to suppress seam loss with a corresponding $Q_{\mathrm{seam}} > 10^9 $ for a resonance frequency of 6~GHz.
	The trenches are machined perpendicular to the coaxial cavities and are used to hold the qubit chips.
	They are aligned to the top of the inner conductor to maximize the coupling strength between the transmon qubit and the coaxial cavity.
	Before assembling, the whole package is etched for four hours with a specific Al etchant to suppress the potential losses due to the oxide and the roughness of the bulk surface~\cite{Reagor2013APL}.
	
	There are two types of designs of transmon qubits: one with a $Y$-shaped antenna and the other with an $I$-shaped antenna.
	Each $Y$-shaped transmon qubit, used as an auxiliary qubit to help the generation and distribution of entanglement, couples to three modes simultaneously, i.e. a logical qubit coaxial cavity, a common coaxial cavity, and a readout resonator. Each $I$-shaped transmon qubit, used as a control qubit to enable the independent error corrections and measurements for the logical qubit, couples to two modes simultaneously, i.e. a logical qubit coaxial cavity and a readout resonator. The qubit chip is loaded into the bulk aluminum through the machined trench and the antennas are aligned with the electrical field of the coaxial cavity to enhance the coupling strength between them.
	All the ends of the antennas are designed to be circular to maximize the coupling strength.
	On each qubit chip, the readout resonator and the Purcell filter are patterned together with the transmon antennas in a single photolithography step~\cite{Axline2016APL}.
	They are designed with a meandering shape to save chip space.
	The Purcell filter is aimed at avoiding the qubit lifetime being limited by the strongly coupled readout resonator.
	
	\subsection{Tantalum transmon qubit}
	The energy relaxation time $T_1$ of the transmon qubit is mainly limited by dielectric loss~\cite{Wang2015APL}.
	Specifically, the two-level systems (TLS) resident in various forms of surface metal oxides contribute largely to the decoherence process.
	%Traditionally, the metals used to fabricate superconducting qubit are Niobium or Aluminum. These metals, especially Niobium, will produce multiple forms of oxides on their surface, leading to the significant presence of the TLSs.
	Tantalum has been recently demonstrated to be a promising material for superconducting qubits~\cite{Place2021NC}.
	The energy relaxation time $T_1$ of transmon qubits made with tantalum films on sapphire substrates has been shown to be over $500~\mu$s~\cite{Wang2022NPJ}.
	
	The qubit chips used in our experiment are fabricated in the laboratory of the Beijing Academy of Quantum Information Sciences ({BAQIS})~\cite{Wang2022NPJ}.
	The coherence time has been greatly improved compared with the traditional Al-only transmon qubit.
	The fabrication procedure contains the following six steps:
	(1) A two-inch sapphire substrate, which can hold seven qubit chips, is cleaned with optimized solvent and piranha solution, and then annealed at temperatures up to 1100°C.
	(2) A 120-nm tantalum film is deposited on the substrate by using dc magnetron sputtering.
	(3) Photolithography and inductively coupled plasma (ICP) are used to define larger structures, such as transmon electrode pads, stripline resonators, and Purcell filters.
	(4) Electron-beam lithography with a resist of PMMA A4/LOR10B is applied to define the junction patterns.
	(5) \textit{In situ} radio frequency ion etching is used followed by a double-angle electron-beam evaporation to prepare the $\mathrm{Al/AlO_{x}/Al}$ Josephson junctions.
	(6) Finally, the wafer is diced into dimensions of 5.5~mm$\times$27.3~mm and mounted into an Al package for low-temperature measurements.
	Between most of the above steps, chemical cleaning is carefully applied to remove the contaminants introduced.

	\begin{table}
		\centering
		\caption{The measured Kerr terms ($ \chi_{ij}/2 \pi $ in MHz) in the system Hamiltonian. \red{$R_X$ represents the readout cavity for qubit ``$X$".}
			The self-Kerr terms for $S_1$, $S_2$, and $S_3$ here are \red{for the situation} where both the adjacent qubits are in their ground states.
			(\textcolor{red}{Note} : --- means the term is not measurable; $ \sim $ means this term is not measured.)}
		\label{tab:SM_Hamiltonian}
		\resizebox{\linewidth}{!}{%{\tiny {\Large {\Large {\tiny }}}}
			\begin{tabular}{>{\centering\hspace{0pt}}m{0.08\linewidth}|>{\centering\hspace{0pt}}m{0.081\linewidth}>{\centering\hspace{0pt}}m{0.081\linewidth}>{\centering\hspace{0pt}}m{0.14\linewidth}>{\centering\hspace{0pt}}m{0.081\linewidth}>{\centering\hspace{0pt}}m{0.081\linewidth}>{\centering\hspace{0pt}}m{0.14\linewidth}>{\centering\hspace{0pt}}m{0.081\linewidth}>{\centering\hspace{0pt}}m{0.081\linewidth}>{\centering\hspace{0pt}}m{0.14\linewidth}>{\centering\hspace{0pt}}m{0.081\linewidth}>{\centering\arraybackslash\hspace{0pt}}m{0.081\linewidth}} 
				\hline
				Mode  & ${R_{{I_1}}}$ & $I_1$  & $S_1$    & $Y_1$  & ${R_{{Y_1}}}$ & $S_2$    & ${R_{{Y_2}}}$ & $Y_2$             & $S_3$    & $I_2$             & ${R_{{I_2}}}$  \\ 
				\hline
				$I_1$ & 3.380         & $\sim$ & 1.300    & 0.003  & ---           & ---      & ---           & 0.005             & ---      & ---               & ---            \\
				$S_1$ & ---           & 1.300  & 0.003    & 0.623  & ---           & $<0.001$ & ---           & ---               & ---      & ---               & ---            \\
				$Y_1$ & ---           & 0.003  & 0.623    & $\sim$ & 3.400         & 1.309    & ---           & 0.019             & ---      & ---               & ---            \\
				$S_2$ & ---           & ---    & $<0.001$ & 1.309  & ---           & 0.004    & ---           & 0.661             & $<0.001$ & ---               & ---            \\
				$Y_2$ & ---           & 0.005  & ---      & 0.019  & ---           & 0.661    & 2.300         & $\sim$            & 0.451    & 0.027             & ---            \\
				$S_3$ & ---           & ---    & ---      & ---    & ---           & $<0.001$ & ---           & 0.451             & 0.002    & 1.411             & ---            \\
				$I_2$ & ---           & ---    & ---      & ---    & ---           & ---      & ---           & 0.027             & 1.411    & $\sim$            & 3.500          \\
				\hline
			\end{tabular}
		}
	\end{table}
	
	\subsection{System Hamiltonian and device parameters}
	
	There are totally fifteen modes in the whole system and they can be roughly divided into two categories, i.e. the nearly-linear modes and the non-linear modes. The nearly-linear modes include the three storage modes, the four readout modes, and the four Purcell filter modes.
	The non-linear modes are the four transmon modes, which have large anharmonicities so that they are treated here as two-level qubits. All the transmon qubits dispersively couple to their adjacent 3D coaxial cavities and the corresponding readout cavities.
	Since the filter modes are never driven directly, they almost stay in their ground states and the associated Hamiltonian terms can be ignored safely.
	Thus the dynamics of the whole system can be described with the following Hamiltonian
	\begin{equation}
		\begin{aligned}
			H/\hbar = &\sum\limits_{i = 1}^4 {{\omega _{{\mathrm{r}}i}}\hat a_{{\mathrm{r}}i}^\dag {\hat a_{{\mathrm{r}}i}}}  + \sum\limits_{i = 1}^4 {{\omega _{{\mathrm{q}}i}}\left| {{e_i}} \right\rangle \left\langle {{e_i}} \right|}  + \sum\limits_{i = 1}^3 {{\omega _{{\mathrm{s}}i}}\hat a_{{\mathrm{s}}i}^\dag {\hat a_{{\mathrm{s}}i}}} \\
			&- \sum\limits_{i = 1}^4 {{\chi _{{\mathrm{r}}i{\mathrm{q}}i}}\left| {{e_i}} \right\rangle \left\langle {{e_i}} \right|\hat a_{{\mathrm{r}}i}^\dag {\hat a_{{\mathrm{r}}i}}} \\
			&- {\chi _{{\mathrm{s}}1{\mathrm{q}}1}}\left| {{e_1}} \right\rangle \left\langle {{e_1}} \right|\hat a_{{\mathrm{s}}1}^\dag {\hat a_{{\mathrm{s}}1}} - {\chi _{{\mathrm{s}}1{\mathrm{q2}}}}\left| {{e_2}} \right\rangle \left\langle {{e_2}} \right|\hat a_{{\mathrm{s}}1}^\dag {\hat a_{{\mathrm{s}}1}}\\
			&- {\chi _{{\mathrm{s2q2}}}}\left| {{e_2}} \right\rangle \left\langle {{e_2}} \right|\hat a_{{\mathrm{s2}}}^\dag {\hat a_{{\mathrm{s}}2}} - {\chi _{{\mathrm{s}}2{\mathrm{q}}3}}\left| {{e_3}} \right\rangle \left\langle {{e_3}} \right|\hat a_{{\mathrm{s}}2}^\dag {\hat a_{{\mathrm{s2}}}}\\
			&- {\chi _{{\mathrm{s}}3{\mathrm{q3}}}}\left| {{e_3}} \right\rangle \left\langle {{e_3}} \right|\hat a_{{\mathrm{s3}}}^\dag {\hat a_{{\mathrm{s3}}}} - {\chi _{{\mathrm{s3q4}}}}\left| {{e_4}} \right\rangle \left\langle {{e_4}} \right|\hat a_{{\mathrm{s3}}}^\dag {\hat a_{{\mathrm{s3}}}}\\
			&- \sum\limits_{i = 1}^3 {\frac{{{K_{{\mathrm{s}}i}}}}{2}} \hat a_{{\mathrm{s}}i}^\dag \hat a_{{\mathrm{s}}i}^\dag {\hat a_{{\mathrm{s}}i}}{\hat a_{{\mathrm{s}}i}}\\
			&- \sum\limits_{i,j = 1\hfill\atop
				i \ne j\hfill}^4 {{\chi _{{\mathrm{q}}i{\mathrm{q}}j}} | {{e_i}} \rangle \langle {{e_i}} |\otimes| {{e_j}} \rangle \langle {{e_j}}|}\\
			&- \sum\limits_{i,j = 1\hfill\atop
				i \ne j\hfill}^3 {{\chi _{{\mathrm{s}}i{\mathrm{s}}j}}\hat a_{{\mathrm{s}}i}^\dag {\hat a_{{\mathrm{s}}i}}\hat a_{{\mathrm{s}}j}^\dag {\hat a_{{\mathrm{s}}j}}}, 
		\end{aligned}
	\end{equation}
	where ${{\mathrm{q}}_{i = 1,2,3,4}} = \{ {I_1},{Y_1},{Y_2},{I_2}\} $ denotes the transmon qubits, $ \omega_{{\mathrm{r}}i} $ is the frequency of the $i$-th readout mode $ {\hat a_{{\mathrm{r}}i}} $, $ \omega_{{\mathrm{q}}i} $ is the transition frequency of the $ i $-th transmon qubit mode between its ground and the first excited states, $\ket{g_i}$ ($\ket{e_i}$) is the ground (the excited) state of the $i$-th transmon qubit, $\omega_{{\mathrm{s}}i} $ is the frequency of the $ i $-th coaxial cavity mode $ {\hat a_{{\mathrm{s}}i}} $, $ {\chi _{{\mathrm{r}}i{\mathrm{q}}i}} $ is the dispersive coupling strength between mode $ {\hat a_{{\mathrm{r}}i}} $ and the $ i $-th transmon qubit, $ {\chi _{{\mathrm{s}}i{\mathrm{q}}j}} $ is the dispersive coupling between mode $ {\hat a_{{\mathrm{s}}i}} $ and the $ j $-th transmon qubit, $ {{{K_{{\mathrm{s}}i}}}} $ is the self-Kerr of the $ i $-th coaxial cavity mode $ {\hat a_{{\mathrm{s}}i}} $, $ {\chi _{{\mathrm{q}}i{\mathrm{q}}j}} $ is the cross-Kerr interaction between the $ i $-th and the $ j $-th transmon qubits, and $ {\chi _{{\mathrm{s}}i{\mathrm{s}}j}} $ is the cross-Kerr interaction between the $ i $-th and the $ j $-th coaxial cavity modes.
	All the above Hamiltonian parameters are listed in Table~\ref{tab:SM_Hamiltonian} and the second column of Table~\ref{tab:SM_coherence}.
	The coherence times and the thermal populations for different modes are measured experimentally and provided in the remaining columns of Table~\ref{tab:SM_coherence}.

	\begin{table}
		\centering
		\caption{The measured mode frequencies, coherence times (the energy relaxation time $T_1$ and \red{the Ramsey dephasing time $T_2^*$}), and thermal populations $n_{\mathrm{th}}$.}
		\label{tab:SM_coherence}
		\begin{tabular}{c|cccc} 
			\hline
			Mode          & Frequency (GHz) & ${T_1}{\text{ (}}\mu $s${\text{)}}$ & $T_2^*{\text{ (}}\mu$s${\text{)}}$ & ${n_{\mathrm{th}}}$ \\ 
			\hline
			${R_{{I_1}}}$ & 8.558           & 0.064                             & ---                               & ---        \\
			$I_1$         & 4.209           & 90-120                               & 40-80                             & 5$\%$       \\
			$S_1$         & 6.102           & 265                               & ---                               & 3$\%$ 			\\
			$Y_1$         & 4.808           & 55-80                                & 60-75                                & 2$\%$       \\
			${R_{{Y_1}}}$ & 8.485           & 0.082                             & ---                               & 1$\%$          \\
			$S_2$         & 5.561           & 300                               & ---                               & 2$\%$      \\
			${R_{{Y_2}}}$ & 8.515           & 0.065                             & ---                               & ---      \\
			$Y_2$         & 4.461           & 100-140                           & 100-140                               & 3$\%$  \\
			$S_3$         & 6.005           & 314                               & ---                               & 2.5$\%$      \\
			$I_2$         & 4.215           & 80-100                            & 30-80                             & 7$\%$   \\
			${R_{{I_2}}}$ & 8.575           & 0.050                             & ---                               & ---     \\
			\hline
		\end{tabular}
	\end{table}
	
	%\subsection{(o) Tab.I. Hamiltonian}
	
	%\subsection{(o) Tab.II. frequency and coherence}
	
	%\section{Experimental setup}
	
	%\subsection{(o) Ta transmon qubit with long lifetime(main or SI?)}
	
	\subsection{The detailed wiring}
	
\begin{figure*}
		\centering
		\includegraphics[scale=1]{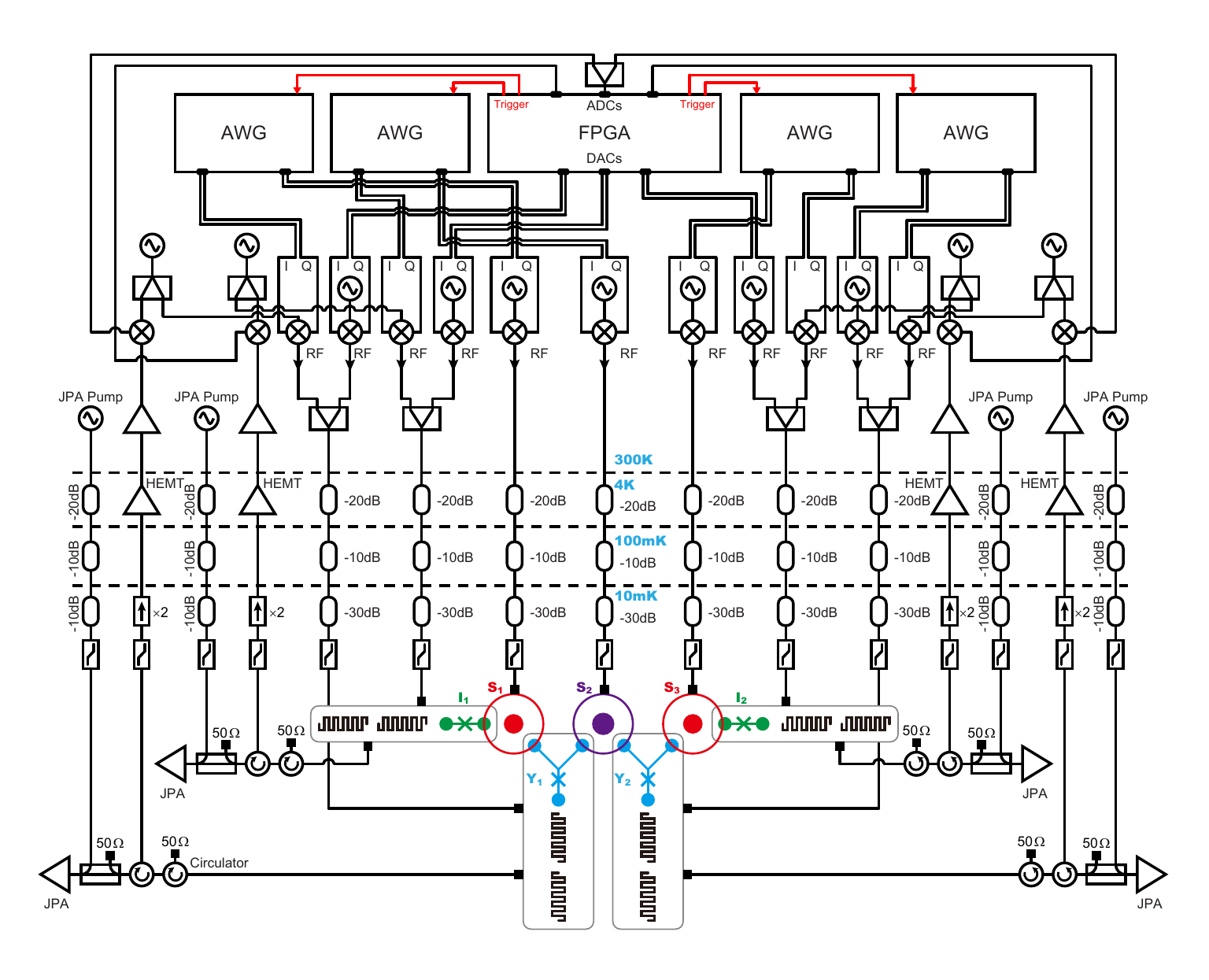}
		\caption[The details for the system wiring]{\textbf{\textcolor{red}{Detailed system wiring.}} 
			The input control signals are generated by the single-sideband technique.
			Then they are attenuated and filtered at the different stages of the refrigerator before going into the experimental device.
			At room temperature, the carrier signal is produced by a microwave generator and the IQ signal for modulation is generated by either AWGs or the DACs of an FPGA.
			The AWGs are triggered by the FPGA DIO ports, to keep all the control sequences synchronized.
			After the IQ mixer, there is also a fast switch for preventing leakage from the carrier signal.
			The measurement signal from the readout cavity is first sent through two circulators and then a JPA for enabling the high-fidelity single-shot readout of the qubit state.
			The pump for the JPA is offered by an independent microwave generator.
			There are also two isolators in the output chain to prevent the back-propagation of the noise.
			The output signal is further amplified with a HEMT amplifier on the 4K stage and another low-noise commercial amplifier at room temperature.
			Before the digitizing of the ADCs, the signal is down-converted to 50~MHz and filtered with a bandpass filter.}
		\label{fig:SM_Wiring}
	\end{figure*}

	The detailed wiring of the whole microwave setup is shown in Fig.~\ref{fig:SM_Wiring}.
	The assembled device is mounted on the base plate of a dilution refrigerator with a temperature of about 10~mK.
	To further reduce the effect of the external magnetic field, the whole sample is enclosed in a box of high-$ \mu $ metal.
	Along the input lines, microwave attenuators and low-pass filters are used to suppress the noise from the higher-temperature plates.
	In this experiment, we use four Josephson parametric amplifiers (JPAs)~\cite{Roy2015APL,Kamal2009PRB} to assist the readout of the qubit states and each JPA is connected to the output port of the readout cavity through two concatenated circulators.
	The JPAs are biased at working points with a gain of about 20~dB and a bandwidth of \red{a few tens~MHz}, allowing separate and high-fidelity single-shot readout of the qubit states.
	%The use of the circulator is to ensure the output signal is sent only in one direction.
	
	All the modes, including the four qubit modes, the three storage cavity modes, and the four readout cavity modes, are controlled by modulated signals.
	The modulation is realized with the assistance of an IQ mixer by using the single-sideband technique.
	For each control signal, there is a local generator that is used to produce the GHz carrier frequency and a pair of IQ signals that are produced by either arbitrary waveform generators (AWGs) or the digital-to-analog converters (DACs) of field programmable gate array (FPGA) cards.
	The IQ signals are about one hundred MHz, whose phase and amplitude are easy to control with high precision.
	The FPGA board itself is capable of performing the full control and readout tasks, i.e. outputting the control pulse, recording the readout signal, and performing feedback operations in real-time. Because of the scale of the experimental device, a total of eleven IQ pairs are needed to run the experimental protocol. We use three FPGA boards to control \red{$I_1$, $I_2$, and $Y_1$}, and four extra AWG5014s to provide the remaining IQ pairs.
	The synchronization of the controls is realized by triggering the AWGs with the digital outputs of the FPGA boards.
	
	The model of the FPGA board is X6-1000M from Innovative Integration and  all three boards are mounted in a single VPXI-ePC chassis.
	The integrated chip on each board is the Xilinx VIRTEX-6 FPGA, with the homemade logic fired in it.
	The sampling rate is controlled by a VPX-COMEX module and we use a 1 GHz clock to operate the ADCs, DACs, and DIOs on the boards.
	The readout signal from the device is first down-converted to 50 MHz and then fed into the ADCs of the FPGA board, where demodulation is carried out and the qubit state is determined as a digitized value of 0 or 1, indicating $ \left| g \right\rangle $ or $ \left| e \right\rangle $ respectively.
	In the qubit measurement and reset process, these 0s or 1s are distributed among the boards through DIO ports as signals for feeding back the qubit reset pulse.
	The total feedback latency, from finishing sending the readout pulse to starting sending the feedback pulse, is 336~ns for our setup \red{including the travel time in the experimental circuitry.}
	
	\section{Experimental characterization}
	
	%\subsection{(o) readout property of each qubit?}
	
	%\begin{figure}
	%	\centering
	%	\includegraphics[scale=1]{SI_ReadoutFidelity.pdf}
	%	\caption{\textbf{The readout fidelity of different qubits in the experiment.}  }
	%	\label{fig:SI_ReadoutFidelity}
	%\end{figure}
	
	\subsection{Readout property of the transmon qubits}
	A good readout performance is of great importance for measurement-based quantum error correction (QEC).
	Although here we adopt the autonomous QEC (AQEC), we still use measurement and feedback to reset the control qubits, and hence the readout performance will largely affect the overall error correction fidelity.
	% We characterize the readout performance by measuring the QND readout fidelity, namely $\mathrm{F}_{gg}$ and $\mathrm{F}_{ee}$ for them.
	% These two quantities indicate to which extent the measured qubit will keep staying in the eigenstates of the measurement operators.
	% $\mathrm{F}_{gg}$($\mathrm{F}_{ee}$) is the probabilities that the qubit is measured to be $\left| g \right\rangle$($\left| e \right\rangle$) state, conditioned on the qubit is initially prepared in the $\left| g \right\rangle$($\left| e \right\rangle$) state.
	% \textcolor{red}{In experiment, we obtain them by first preparing each qubit in their superposition state of $\left( {\left| g \right\rangle  + \left| e \right\rangle } \right)/2$ and then apply two successive measurements.
		% The first measurement is used as preparing the initial state by post-selection, and the second measurement result is used to extract the corresponding $\mathrm{F}_{gg}$($\mathrm{F}_{ee}$).}
	In the experiment, the transmon qubit is first measured and the ground state $\left| g \right\rangle$  is post-selected, and then no $\pi$ pulse or a $\pi$ pulse is implemented to prepare the $\left| g \right\rangle $ or $\left| e \right\rangle $ state, followed immediately by a second readout measurement.
	The readout fidelities $F_\mathrm{g}$ and $F_\mathrm{e}$ are inferred from the histograms of the second readout in the two cases.
	The results for all the transmon qubits are listed in Table~\ref{tab:SM_readout_Fidelity}.
	Note that the readout infidelity for ${\left| e \right\rangle }$ is mainly due to the qubit damping effect in the readout and the following waiting time.
	
	\textbf{Data analysis}. Due to the mismatch between the dispersive shift and the damping rate of the readout cavity, the signal-to-noise ratio is not optimal and the two readout Gaussian blobs are not completely separated. In addition, the qubit damping effect in the readout process degrades the overall readout performance further.
	To mitigate these two factors and infer the qubit state more accurately, we adopt a calibration matrix based on Bayes's rule, which is constructed by
	\begin{equation}
		{F_{{\mathrm{Bayes}}}} = \left( {\begin{array}{*{20}{c}}
				{{F_{\mathrm{g}}}}&{1 - {F_{\mathrm{e}}}}\\
				{1 - {F_{\mathrm{g}}}}&{{F_{\mathrm{e}}}}
		\end{array}} \right).
	\end{equation}
	Then the final qubit readout result $ {P_f} $ can be expressed as
	\begin{equation}
		{P_f} = F_\mathrm{Bayes}^{ - 1} \cdot {P_m},
	\end{equation}
	where $ {P_m} $ is the uncorrected measurement result.
	The Bayes calibration matrix of each qubit is calibrated separately.
	
	\textbf{$\pi$ pulse fidelity during AQEC}. In the experiment, a reset pulse dependent on the measurement result is implemented after each AQEC pulse. Such a control pulse has a Gaussian envelope function with $\sigma=5$~ns at the qubit transition frequency corresponding to its coupled cavity at Fock state $\ket{2}$. The fidelity of such a reset pulse is 0.9844 and 0.9838 for qubits $I_1$ and $I_2$, respectively. The infidelity mainly comes from the reset pulse being off-resonant for the qubit when the coupled cavity is at Fock state $\ket{0}$ or $\ket{4}$.
	%(**prefer the former expression**\red{The infidelity mainly comes from the non-perfect amplitude calibration, finite bandwidth of the pulses, or possible calibration fluctuations.})
	
	\begin{table}
		\begin{tabular}{p{1.5cm}<{\centering}p{1.5cm}<{\centering}p{1.5cm}<{\centering}p{1.5cm}<{\centering}p{1.5cm}<{\centering}}
			\hline
			\centering
			& $I_1$ & $I_2$ & $Y_1$ & $Y_2$\\
			\hline
			$F_\mathrm{g}$  & 0.9918     & 0.9829      & 0.9984     & 0.9937   \\
			$F_\mathrm{e}$   & 0.9896     & 0.9784      & 0.9926    & 0.9926   \\
			\hline
		\end{tabular} \vspace{8pt}\\
		\caption{\textbf{The readout fidelities of all four qubits in the experiment.} $F_\mathrm{g}$ ($F_\mathrm{e}$) corresponds to the probability of being measured to be $\left| g \right\rangle$($\left| e \right\rangle$), conditional on applying no or a $\pi$ pulse on the initially post-selected ground state. The slightly lower value of $F_\mathrm{e}$ for each qubit is mainly due to the decoherence of the $\left| e \right\rangle$ state in the process of readout and waiting time.}
		\label{tab:SM_readout_Fidelity}
	\end{table}
	
	\subsection{Parity measurement fidelity}
	In our experiment, the photon-number parity serves as the error syndrome for error detection. The parity measurement is also used in joint-Wigner tomography and cavity state initialization. There are two protocols that can be used to implement the parity measurement by mapping different parities onto the different states of the control qubit. They both take the form of a Ramsey-like pulse sequence, $ (R_{\pi /2}^Y,\pi /{\chi _{\mathrm{sq}}},R_{\pi /2}^Y) $ and $  (R_{\pi /2}^Y,\pi /{\chi_{\mathrm{sq}}},R_{ - \pi /2}^Y) $, where $ R_{\pm \pi /2}^Y $ is the unconditional qubit rotation of $ {\pm \pi /2} $ around the y axis and $ \pi /{\chi _{\mathrm{sq}}} $ is the free evolution time between the two qubit pulses~\cite{Sun2014Nature} with $\chi_{\mathrm{sq}}$ being the dispersive coupling strength.
	The only difference is that the first protocol maps the even cavity state onto $ \left| e \right\rangle $ and the odd one onto $ \left| g \right\rangle $, while the second protocol does the opposite.
	
	%Taking the second protocol $  (R_{\pi /2}^Y,\pi /{\chi_{\mathrm{qs}}},R_{ - \pi /2}^Y) $ as an example, by starting with the even cat state of $\left| g \right\rangle \frac{1}{{\sqrt 2 }}\left( {\left| \alpha  \right\rangle  + \left| { - \alpha } \right\rangle } \right)$, the first $R_{\pi / 2}^{Y}$ rotation brings the state into $\frac{1}{2}(|g\rangle+|e\rangle)(|\alpha\rangle+|-\alpha\rangle)$. After free evolution time of $t=\pi/\chi_{\mathrm{qs}}$ under the dispersive interaction, the state becomes $\frac{1}{2}|g\rangle(|\alpha\rangle+|-\alpha\rangle)+\frac{1}{2}|e\rangle(|-\alpha\rangle+|\alpha\rangle)$, with the cavity phase being rotated with $\pi$. Then the second $R_{-\pi / 2}^{Y}$ will end mapping the parity information onto the transmon qubit state, $|g\rangle \frac{1}{\sqrt{2}}(|\alpha\rangle+|-\alpha\rangle)$.
	In our experiment, we choose the second protocol to carry out the parity measurement, because the logical qubit encoding with the lowest binomial code has an even parity. For a low probability of error with the current correction cycle length, mapping even parity to $|g\rangle$ has a higher parity measurement fidelity.
	
	We obtain the average parity measurement fidelity by first spreparing a coherent state with a displacement operation $D(\alpha=\sqrt{\bar{n}})$  with $\bar{n}=0,1,2,3$, and then carry out three consecutive parity measurements using the second protocol $(R_{\pi /2}^Y,\pi /{\chi_{\mathrm{sq}}},R_{ - \pi/2}^Y)$.
	The varying $\bar{n}$ is used to explore the dependence of fidelity on the average photon number in the storage cavity, because for a larger $\bar{n}$ the qubit $R_{\pm \pi /2}^Y$ rotation becomes conditional on the photon number due to the finite bandwidth of the rotation pulses.
	The first two parity measurements are aimed at obtaining an ideal even or odd parity state by post-selecting the same results. The last measurement gives the fidelity value for the even or odd state separately, and the average parity measurement fidelity is obtained by taking the mean of these two values.
	The parity measurement fidelities for $S_1$ and $S_3$ are summarized in Table~\ref{tab:SM_parity_fidelity}.
	The most relevant values are those for $\bar{n}=2$ which is the average photon number for the logical bases.
	The infidelity in this parity measurement process mainly comes from two aspects: the qubit decoherence error during the free evolution time and the final readout error after the mapping process.
	
	\begin{table}
		\caption{\textbf{Average parity measurement fidelity vs  $\Bar{n}$  based on the
				second protocol $(R_{\pi/2}^{Y}, \pi/\chi_{\mathrm{sq}}, R_{-\pi/2}^{Y})$.}}
		\label{tab:SM_parity_fidelity}
		\begin{tabular}{p{1.5cm}<{\centering}p{3cm}<{\centering}p{3cm}<{\centering}}
			\hline
			\centering
			$\Bar{n}$ & $S_1$ parity measurement fidelity & $S_3$ parity measurement fidelity \\
			\hline
			0     & 0.9946      & 0.9930       \\
			1    & 0.9664      & 0.9751     \\
			2    & 0.9629     & 0.9705    \\
			3    & 0.9473      & 0.9632     \\
			\hline
		\end{tabular} \vspace{8pt}\\
	\end{table}
	
	\subsection{Calibration and compensation of measurement-induced phase}
	
	\begin{figure}
		\centering
		\includegraphics[scale=1]{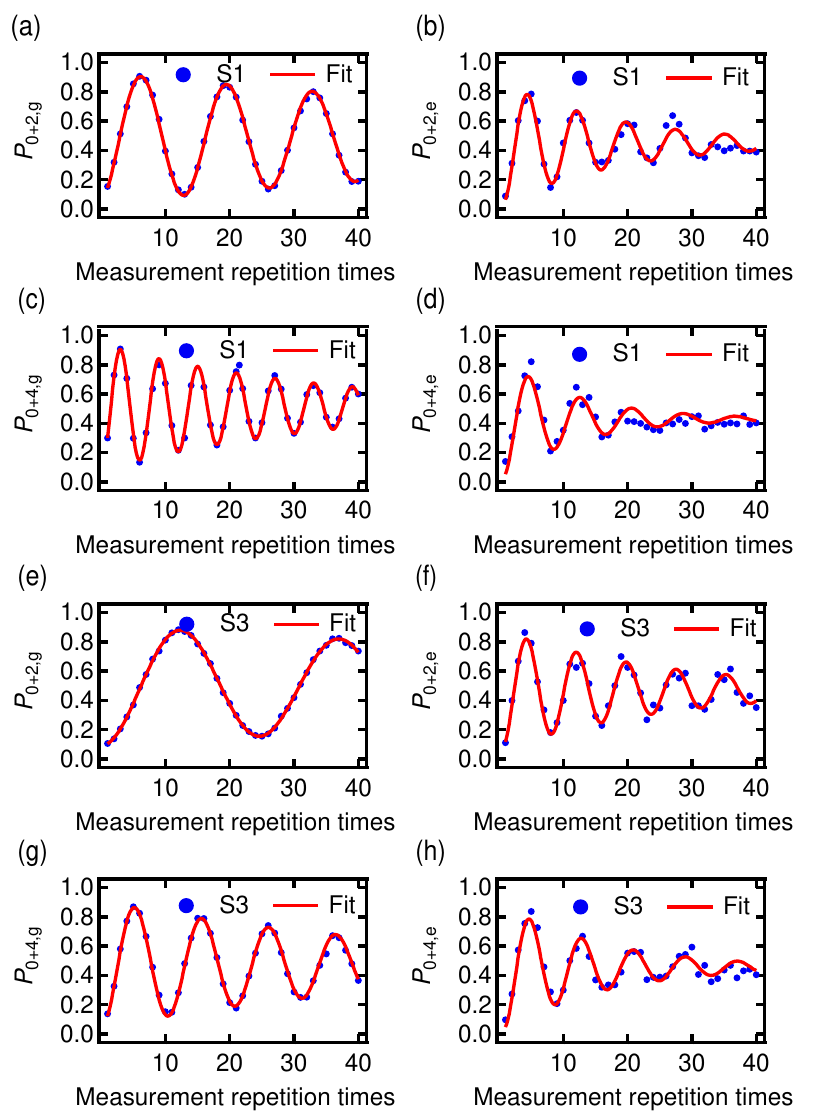}
		\caption{\textbf{Calibration results of the measurement-induced phases.} 
			Only the phase accumulations of Fock states $\left| 2 \right\rangle$ and $\left| 4 \right\rangle$ are calibrated for $S_1$ (a-d) and for $S_3$ (e-h).
			Both phases are conditional on the corresponding control qubit being in the ground or the excited states and are characterized separately.
			The joint system of the storage cavity and the control qubit is first prepared in 
			\red{$\left( {\left| 0 \right\rangle  + \left| N \right\rangle } \right)/\sqrt{2} \otimes \left| g \right\rangle$ or $\left( {\left| 0 \right\rangle  + \left| N \right\rangle } \right)/\sqrt{2} \otimes \left| e \right\rangle$, with $N=2,4$.}
			%\red{$\left( {\left| 0 \right\rangle  + \left| 2 \right\rangle } \right)/\sqrt{2} \otimes \left| g \right\rangle$, $\left( {\left| 0 \right\rangle  + \left| 2 \right\rangle } \right)/\sqrt{2} \otimes \left| e \right\rangle$, $\left( {\left| 0 \right\rangle  + \left| 4 \right\rangle } \right)/\sqrt{2} \otimes \left| g \right\rangle$ and $\left( {\left| 0 \right\rangle  + \left| 4 \right\rangle } \right)/\sqrt{2} \otimes \left| e \right\rangle$ respectively.}
			\red{Then the probability $P_{0+N,g}$ or $P_{0+N,e}$ of the storage cavity conditioned on the control qubit state along the corresponding basis of $\left( {\left| 0 \right\rangle  + \left| N \right\rangle } \right)/\sqrt{2}$ is detected after repeating the measurement operation 1 to 40 times.}
			%\red{$P_{0+N,g}$ of measuring $(\ket{0}+\ket{N})/\sqrt{2}\otimes \ket{g}$, or $P_{0+N,e}$ of $(\ket{0}+\ket{N})/\sqrt{2}\otimes \ket{e}$}, 
			The blue dots are the experimental results and the red lines are their fitting curves with a damped sinusoidal function.}
		\label{fig:SM_RIP}
	\end{figure}
	
	The measurement-induced phase is the phase accumulation on the storage state caused by the population in the readout cavity through the cross-Kerr interaction ${\chi _{{\mathrm{sr}}}}\hat a_{\mathrm{s}}^{^\dagger }{\hat a_{\mathrm{s}}}\hat a_{\mathrm{r}}^{^\dagger }{\hat a_{\mathrm{r}}}$.
	In order to recover the originally encoded state, this phase needs to be canceled as part of the AQEC operation.
	We note that during the measurement and the subsequent waiting time, there is also a phase accumulation caused by the self-Kerr effect, $\frac{{{K_{\mathrm{s}}}}}{2}\hat a_{\mathrm{s}}^\dagger \hat a_{\mathrm{s}}^\dagger {\hat a_{\mathrm{s}}}{\hat a_{\mathrm{s}}}$, of the storage cavity itself.
	%The difference between the two phases is that the former is proportional to the photon number in the storage cavity, while the latter is of a second-order relation with that.
	We also note that these values are slightly different when qubit stays in its ground or excited state due to the higher-order non-linearity.
	Here we treat these two phase accumulations together and use the following protocol to calibrate them.

	First, the joint system of the storage cavity and the corresponding control qubit is prepared in the states of $\left( {\left| 0 \right\rangle  + \left| N \right\rangle } \right)/\sqrt{2} \otimes \left| g \right\rangle$ or $\left( {\left| 0 \right\rangle  + \left| N \right\rangle } \right)/\sqrt{2} \otimes \left| e \right\rangle$, with $N=2,4$. 
	%\red{$\left( {\left| 0 \right\rangle  + \left| 2 \right\rangle } \right)/\sqrt{2} \otimes \left| g \right\rangle$, $\left( {\left| 0 \right\rangle  + \left| 2 \right\rangle } \right)/\sqrt{2} \otimes \left| e \right\rangle$, $\left( {\left| 0 \right\rangle  + \left| 4 \right\rangle } \right)/\sqrt{2} \otimes \left| g \right\rangle$ and $\left( {\left| 0 \right\rangle  + \left| 4 \right\rangle } \right)/\sqrt{2} \otimes \left| e \right\rangle$ respectively}.
	Next, variable numbers of measurement operations are carried out to accumulate integer times of the measurement-induced phase.
	Finally, the probability $P_{0+N,g}$ or $P_{0+N,e}$ of the storage cavity conditioned on the control qubit state is measured along the corresponding basis of $\left( {\left| 0 \right\rangle  + \left| N \right\rangle } \right)/\sqrt{2}$. %\red{the conditional probability $P_{0+N,g}$ of measuring $(\ket{0}+\ket{N})/\sqrt{2}\otimes \ket{g}$, or $P_{0+N,e}$ of $(\ket{0}+\ket{N})/\sqrt{2}\otimes \ket{e}$, is determined by mapping the storage state onto the corresponding ancilla qubit and measuring the qubit state.}
	The results for $S_1$ and $S_3$ are presented in Fig.~\ref{fig:SM_RIP}.
	The accumulated phase could be extracted by fitting the data with a damped sinusoidal function.
	In this way, the deterministic phase accumulation can be taken into account in the AQEC pulse and corrected in advance.
	
	\begin{table}
		\caption{\textbf{The measurement-induced phases for $S_1$ and $S_3$ obtained in the experiment.}  }
		\label{tab:SM_RIP}
		\begin{tabular}{p{4cm}<{\centering}p{4cm}<{\centering}}
			\hline
			\centering
			&  measurement-induced phase (rad) \\
			\hline
			$|2\rangle$ in $S_1$ with  $|g\rangle$ of $I_1$  &  -0.4712  \\
			$|4\rangle$ in $S_1$ with  $|g\rangle$ of $I_1$  &  -1.0444   \\
			$|2\rangle$ in $S_1$ with $|e\rangle$ of $I_1$ &   -0.2067 \\
			$|4\rangle$ in $S_1$ with $|e\rangle$ of $I_1$ &   -0.4837  \\
			\hline
			$|2\rangle$ in $S_3$ with $|g\rangle$ of $I_2$  &    -0.2549 \\
			$|4\rangle$ in $S_3$ with $|g\rangle$ of $I_2$  &    -0.6027   \\
			$|2\rangle$ in $S_3$ with $|e\rangle$ of $I_2$  &    -0.3039\\
			$|4\rangle$ in $S_3$ with $|e\rangle$  of $I_2$ &    -0.1277 \\
			\hline
		\end{tabular} \vspace{8pt}\\
	\end{table}
	
	\begin{figure*}
		\centering
		\includegraphics[scale=1]{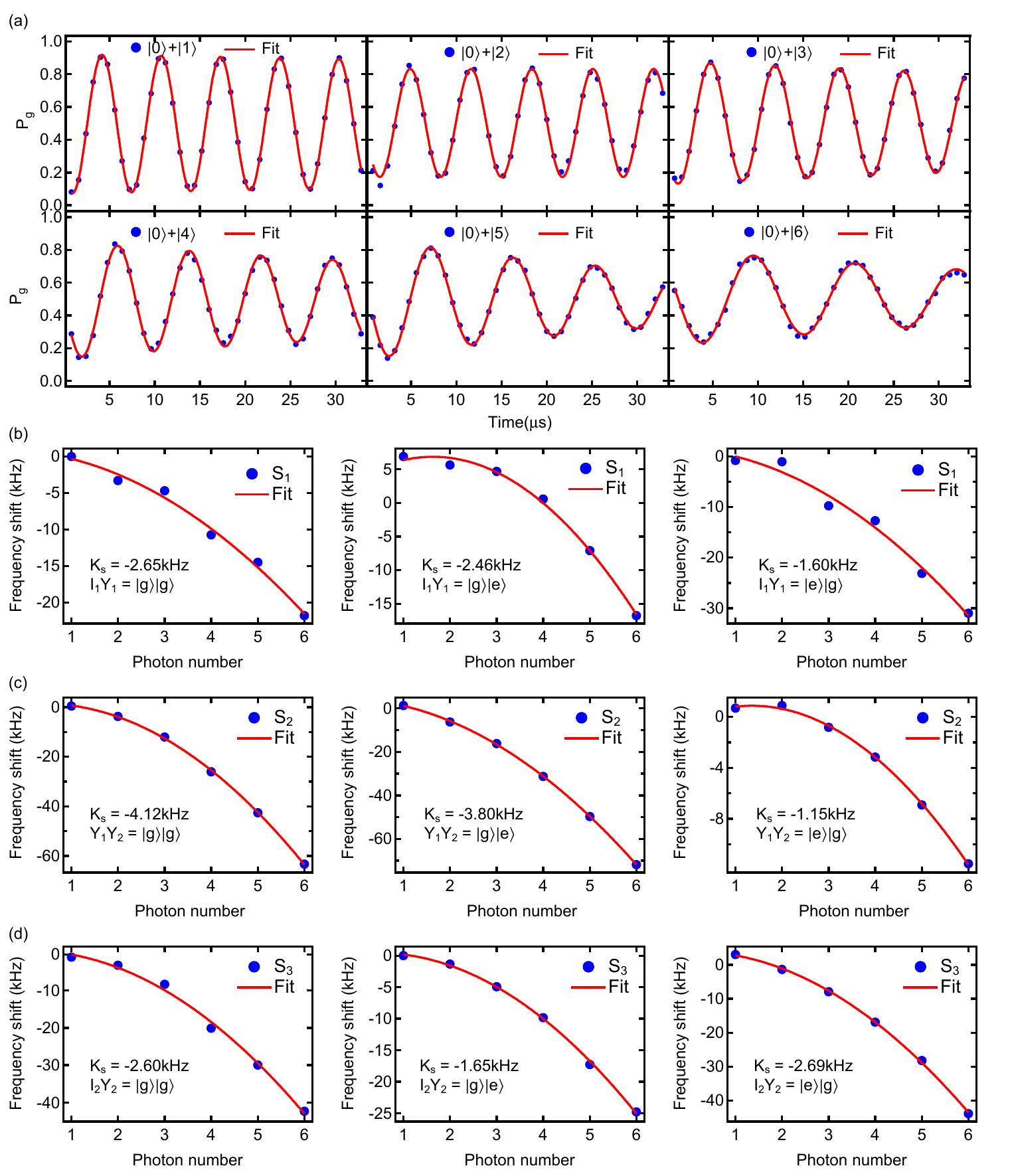}
		\caption{\textbf{The self-Kerr calibrations of the three coaxial cavities.} \red{The calibration is performed separately for each cavity. The cavity is encoded into $(\left| 0 \right\rangle  + \left| N \right\rangle )/\sqrt{2}$, with $N=1,2,...,6$.
				To take into account the higher-order non-linearity, we prepare the adjacent transmon qubits into different combinations of the ground and excited states. The superposition state of $(\left| 0 \right\rangle  + \left| N \right\rangle )/\sqrt{2}$ will acquire relative phases with varying evolution times, and for different $N$ the phase accumulation rate is different. As an example, the calibrations for $S_2$ are shown in (a) \red{with both $Y_1$ and $Y_2$ being in the ground states in the process of cavity phase accumulation}. By fitting with a damped sinusoidal function, $y=y_{0}+A e^{-t / \tau} \cos \left(\omega t+\varphi_{0}\right)$, the rate could be extracted and it is of second-order relationship with the photon number $N$. The results for $S_1$, $S_2$, and $S_3$ are shown in (b), (c), and (d), respectively. Fitting the curves with a quadratic function gives the Kerr coefficients for each cavity in the presence of different adjacent transmon qubit states.}}
		\label{fig:SM_SelfKerr}
	\end{figure*}
	
	\subsection{Self-Kerr coefficients of the storage cavities}
	%\textcolor{red}{Table: Hamiltonian parameters with gg, ge, eg?}
	The self-Kerr effect of the storage cavity can be described by the Hamiltonian ${H_{{\mathrm{Kerr }}}} = - \frac{{{K_{\mathrm{s}}}}}{2}\hat a_{\mathrm{s}}^{\dagger 2}\hat a_{\mathrm{s}}^2$, where $K_\mathrm{s}$ is the self-Kerr coefficient~\cite{Kirchmair2013Nature}.
	This effect will lead to a unitary evolution of the cavity state, where Fock state $\left| n \right\rangle$ acquires a phase $- \frac{{{K_{\mathrm{s}}}}}{2}n(n - 1)t$ with $t$ being the evolution time.
	Since the evolution due to the Kerr effect is unitary, it will not cause the loss of information. However, it is necessary to carefully characterize the Kerr coefficient for precise control of the cavity state.
	
	When taking into account the higher-order correction to the Kerr coefficient, the self-Kerr Hamiltonian can be modified as
	\begin{eqnarray}
		H_{{\mathrm{Kerr }}}/\hbar =  &- \frac{{{K_{\mathrm{s}}}}}{2}\hat a_{\mathrm{s}}^{\dagger 2}\hat a_{\mathrm{s}}^2{\left| g_1 \right\rangle }{\left\langle g_1\right|}\otimes{\left|g_2 \right\rangle}{\left\langle g_2 \right|}\nonumber\\ 
		&+ \frac{\chi_1}{2}\hat a_{\mathrm{s}}^{\dagger 2}\hat a_{\mathrm{s}}^2{\left| g_1 \right\rangle }{\left\langle g_1\right|}\otimes{\left|e_2 \right\rangle}{\left\langle e_2 \right|}\nonumber\\
		&+ \frac{\chi_2}{2}\hat a_{\mathrm{s}}^{\dagger 2}\hat a_{\mathrm{s}}^2{\left| e_1 \right\rangle}{\left\langle e_1 \right|}\otimes{\left|g_2 \right\rangle}{\left\langle g_2 \right|}\nonumber\\
		&+ \frac{\chi_3}{2}\hat a_{\mathrm{s}}^{\dagger 2}\hat a_{\mathrm{s}}^2{\left| e_1 \right\rangle}{\left\langle e_1\right|}\otimes{\left|e_2 \right\rangle}{\left\langle e_2 \right|},
	\end{eqnarray}
	where $\chi_1$, $\chi_2$, and $\chi_3$ are the non-linear corrections when the adjacent qubits are in different states.
	In order to calibrate these items, a superposition state $(\left| 0 \right\rangle  + \left| N \right\rangle )/\sqrt{2}$ with $N=1,2,...,6$ is prepared in the storage cavity while the adjacent qubits are prepared into different combinations of the ground and excited states.
	The joint state will evolve under the above Hamiltonian and accumulate a phase that is quadratic to the photon number $N$ and proportional to the evolving duration $t$ and the Kerr coefficient.
	After variable evolving times, we map the cavity state onto one of the qubits and measure the qubit. Then, we obtain an oscillating curve, from which the corresponding Kerr coefficient can be extracted.
	The calibrations for $S_1$, $S_2$, and $S_3$ are performed separately and the results are shown in Fig.~\ref{fig:SM_SelfKerr}.
	
	\subsection{Qubit-qubit cross-talks}
	
\begin{figure*}
		\centering
		\includegraphics[scale=1]{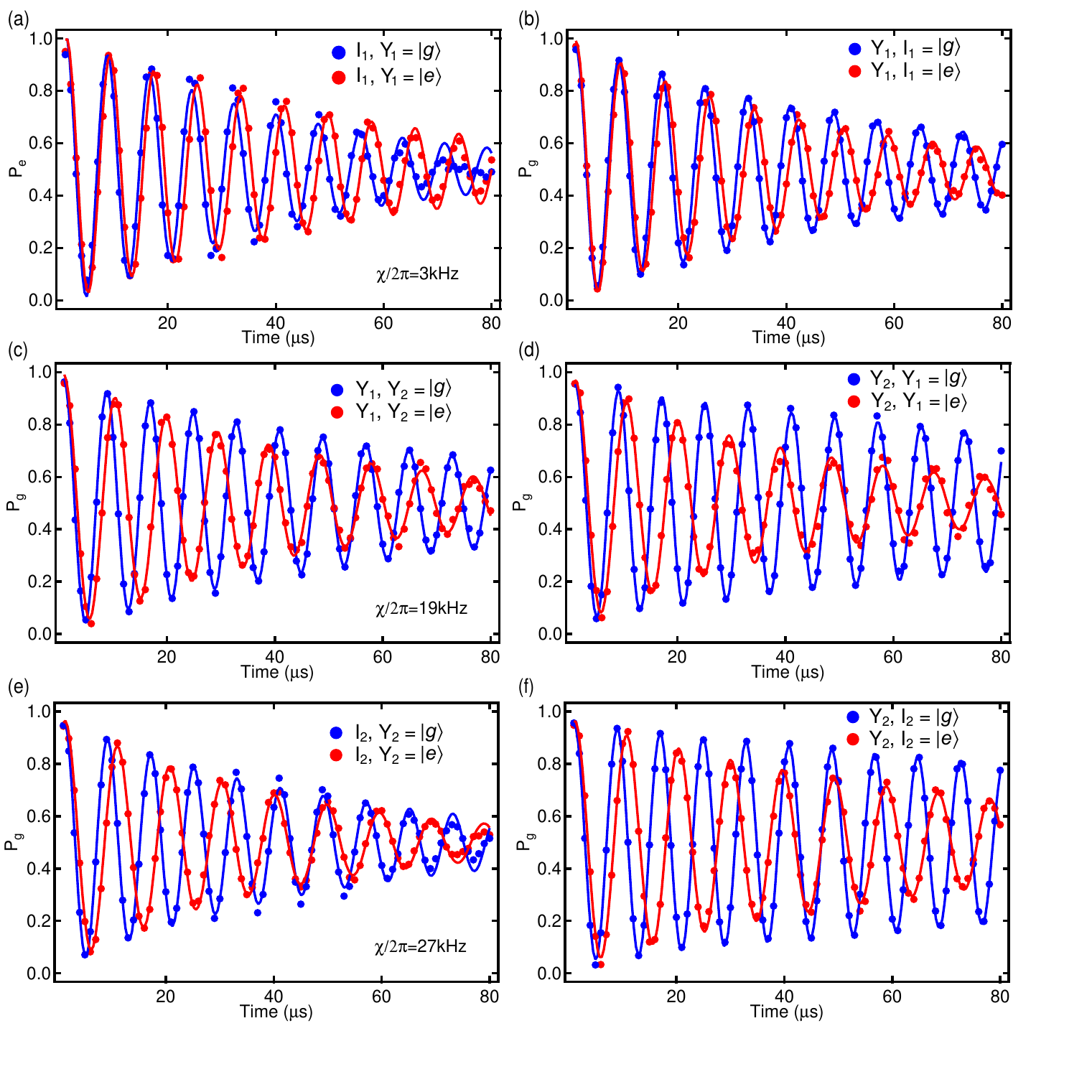}
		\caption{\textbf{The calibration of the transmon qubit cross-talk $\chi_{\mathrm{q}_i \mathrm{q}_j}$.} 
			The direct qubit-qubit coupling term $\chi_{\mathrm{q}_i \mathrm{q}_j}$ is measured for three pairs of adjacent transmon qubits.	The oscillation curve is obtained with a Ramsey experiment on qubit $i$ with qubit $j$ being initialized in the ground or the excited state. For each oscillation, fitting with a damped sinusoidal function gives an oscillation frequency.
			The frequency difference of the blue and red curve gives $\chi_{I_1Y_1} =3$~kHz (a), $\chi_{Y_1Y_2}=19$~kHz (c), and $\chi_{I_2Y_2}=27$~kHz (e).
			We note that the cross-talk between $Y_1$ and $Y_2$ needs to be taken into account when preparing the initial entangled state between $Y_1$ and $Y_2$.
			The other two terms $\chi_{I_1Y_1}$ and $\chi_{I_2Y_2}$ are not that important, because $Y_1$ and $Y_2$ are expected to return to the ground states after the encoding process.}
		\label{fig:SM_qubitCrossTalk.pdf}
	\end{figure*}

	The cross-talk interaction between adjacent qubits can be expressed as
	\begin{equation}
		H_{\mathrm{crosstalk}}/\hbar  =  - {\chi _{{\mathrm{q}}i{\mathrm{q}}j}}|{e_i}\rangle \langle {e_i}| \otimes |{e_j}\rangle \langle {e_j}|,
	\end{equation}
	with ${{\mathrm{q}}_{i = 1,2,3,4}} = \{ {I_1},{Y_1},{Y_2},{I_2}\} $ denoting the qubits and $i \neq j$.
	This means an excitation in the $i$-th qubit will shift the frequency of the $j$-th qubit by an amount of ${\chi _{{\mathrm{q}}i{\mathrm{q}}j}}$, which leads to extra operation errors.
	For this reason, it is essential to take into account the cross-talk between adjacent transmon qubits in the optimized control pulses to avoid undesired accumulated phases.
	Those interactions between the non-adjacent transmon qubits can be safely ignored. The coefficients of cross-talk interaction are calibrated by a Ramsey-like experiment of a transmon qubit while keeping the adjacent transmon qubit in the ground or the excited state. The measured results are shown in Fig.~\ref{fig:SM_qubitCrossTalk.pdf}.
	
	\subsection{Photon-number-resolved ac Stark shift (PASS)}
	
	\begin{figure*}
		\centering
		\includegraphics[scale=1]{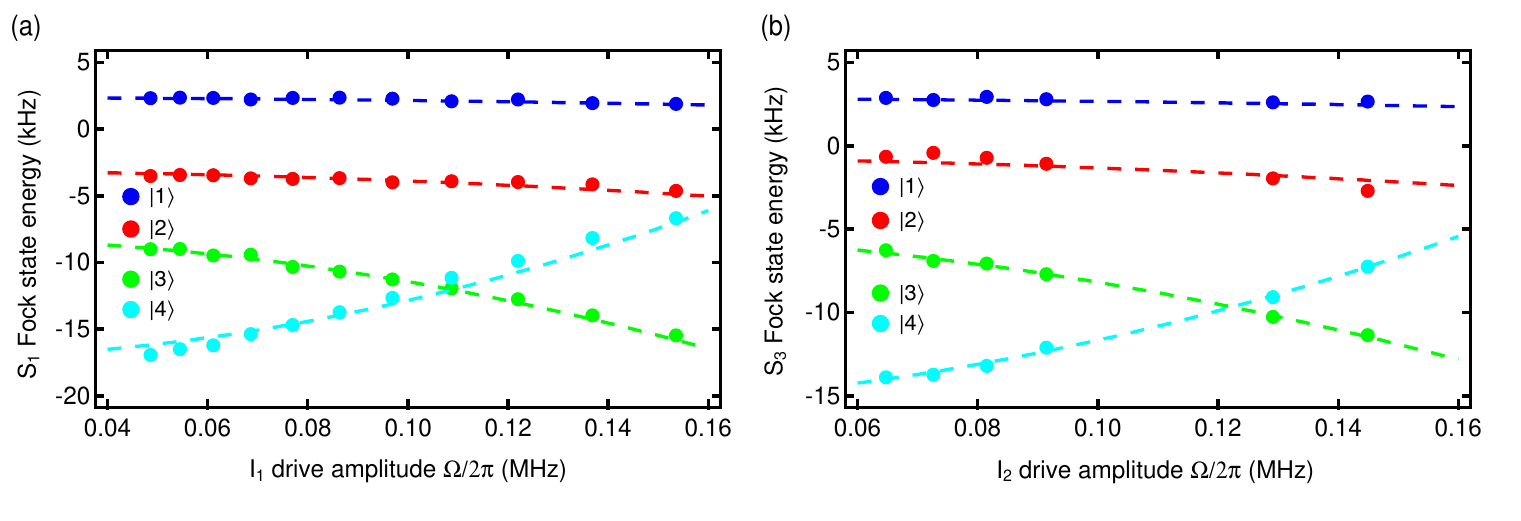}
		\caption{\textbf{The measured Fock state $\left| n \right\rangle$ frequency shift with respect to the amplitude of the detuned drive.} The results are for $S_1$ (a) and $S_3$ (b), respectively.
			The frequency of the detuned drive is chosen to be ${\omega _{\mathrm{d}}} = {\omega _{\mathrm{q}}} - 3.5\chi$, i.e. at the middle of the qubit resonant frequencies corresponding to Fock states $\left| 3 \right\rangle$ and $\left| 4 \right\rangle$.
			As predicted by the theoretical model, the frequency shift $-\frac{\Omega^{2}}{(3.5-n) \chi}$  for different Fock state $\ket{n}$ is either positive or negative. The amplitude strengths that are used to mitigate the photon-jump-induced dephasing of the logical qubits are $0.067\chi_{S_1I_1}$ (0.0863 MHz) and $0.074\chi_{S_3I_2}$  (0.0968 MHz) for qubits $I_1$ and $I_2$, respectively.  }
		\label{fig:SM_PASS}
	\end{figure*}
	
	\begin{table*}
		\caption{\textbf{Frequency shifts under the PASS drives in the experiment.}  }
		\label{tab:SM_PASSdrives}
		\begin{tabular}{p{3.5cm}<{\centering}p{1.5cm}<{\centering}p{1.5cm}<{\centering}p{1.5cm}<{\centering}p{1.5cm}<{\centering}p{1.5cm}<{\centering}p{1.5cm}<{\centering}}
			\hline
			\centering
			Fock state frequency shift (kHz) & $|1\rangle$& $|2\rangle$ & $|3\rangle$ & $|4\rangle$ & $f_3-f_1$ & $f_4-f_2$  \\
			\hline
			$S_1$  & -0.28 & -3.48 & -11.50 & -14.79  & -11.22 & -11.31 \\
			$S_3$  & 0.23 & -2.68 & -10.47 & -13.38  & -10.71 & -10.70    \\
			\hline
		\end{tabular} \vspace{8pt}\\
	\end{table*}
	
	Because of the randomness nature of the photon-loss error and the non-commutativity of the Kerr Hamiltonian $\frac{K}{2} \hat{a}^{\dagger 2} \hat{a}^{2}$ and the annihilation operator $\hat{a}$, the random occurrence time of the photon-loss event will cause a random phase accumulation for the logical qubits, leading to effective dephasing~\cite{Leghtas2013PRL}.
	We address this issue by using the PASS technique (photon-number-resolved ac-Stark shift), where the frequencies of the first four Fock states can be carefully engineered and the phase accumulation can thus be made irrelevant to the error occurrence time~\cite{Ma2020NP}.
	Here we briefly introduce the theoretical model and present our calibration result for the two logical qubits.
	
	For a system of a 3D coaxial cavity and a dispersively coupled transmon qubit, the PASS Hamiltonian can be written as~\cite{Ma2020NP}
	\begin{equation}
		{H_{{\mathrm{PASS}}}/\hbar} = \sum\limits_n {{\delta _n}} \left| n \right\rangle \left\langle n \right| = \sum\limits_n { - \frac{{{\Omega ^2}}}{{{\Delta _{\mathrm{d}}} - n\chi }}\left| n \right\rangle \left\langle n \right|},
	\end{equation}
	where $\Delta_{\mathrm{d}} \equiv \omega_{\mathrm{q}}-\omega_{\mathrm{d}}$ is the detuning of the drive frequency with respect to the qubit transition frequency, $\Omega$ is the Rabi drive frequency of the detuned drive with the condition of $\Omega \ll \left| {{\Delta _{\mathrm{d}}} - \chi {a^\dag }a} \right|$ for all cavity states to ensure the validity of the PASS Hamiltonian, and
	$\chi$ is the dispersive interaction strength between the storage cavity and the transmon qubit. Thus, with the application of a detuned drive, the frequency of the different Fock states can be adjusted precisely.
	
	In the experiment, we characterize the PASS drive by applying one drive with the frequency chosen at ${\Delta _{\mathrm{d}}} = 3.5\chi$ and with varying amplitudes.
	Figure~\ref{fig:SM_PASS} shows the measured Fock state frequency $f_n = - n(n - 1)\frac{K}{2} - \frac{{{\Omega ^2}}}{{{(3.5 - n)\chi }}}$, with respect to an ideal harmonic oscillator with neither the Kerr effect nor the detuned drive.
	The PASS drives for the two 3D coaxial cavities, $S_1$ and $S_3$, are characterized independently.
	From Fig.~\ref{fig:SM_PASS}, it can be seen that the frequency shifts for Fock states $\left| 1 \right\rangle$, $\left| 2 \right\rangle$, and $\left| 3 \right\rangle$ are negative while that for Fock state $\left| 4 \right\rangle$ is positive.
	Thus at a certain drive amplitude, indicated by the dashed line in Fig.~\ref{fig:SM_PASS}, the error-transparent condition ${f_4} - {f_2} = {f_3} - {f_1}$ can be satisfied, where the phase accumulation becomes irrelevant with whether there is a photon-loss error or not. The frequency shifts under the PASS drives in the experiment are shown in Table~\ref{tab:SM_PASSdrives}.
	
	\section{Autonomous quantum error correction (AQEC)}
	\label{sec:Autonomous quantum error correction}
	\subsection{Scheme of the AQEC process}
	Here, we briefly describe the AQEC process for binomial codes. Traditionally, QEC is performed with a two-step process, including error detection and error correction. This is accomplished by measuring the error syndrome operator, here the photon number parity operator, to indicate the occurrence of single-photon-loss errors.
	The following is an adaptive unitary operation $  U_n^\dagger $, which recovers the logical state from the $ n^{\mathrm{th}} $ error space to the original code space.
	The whole process is assisted by an ancilla qubit. % with a Hilbert space of ${{\mathcal H}_A} $ whose dimension is greater than the number of errors in the correctable error set.
	First, we need to apply a unitary gate $ { U_{{\mathrm{map}}}} $ to the qubit and cavity system to map the error in the logical state onto the state of the ancilla:
	\begin{equation}
		\begin{aligned}
			{ U_{{\mathrm{map }}}} = & \sum\limits_n {{{ P}_{{ C_n}}}}  \otimes \left( {\left| n \right\rangle \left\langle 0 \right| + \left| 0 \right\rangle \left\langle n \right| + \sum\limits_{m \ne \{ 0,n\} } {\left| m \right\rangle \left\langle m \right|} } \right)\\ 
			&+ { I_{ H_{\mathrm{O}} - \sum\nolimits_n {{ C_n}} }} \otimes { I_{{ H_{\mathrm{A}}}}}.
		\end{aligned}
	\end{equation}
	Here, $ \left| n \right\rangle $ is the $ n^{\mathrm{th}} $ excitation of the ancilla, ${{{P}_{{ C_n}}}} $ is the projector onto the $ n^{\mathrm{th}} $ error subspace, and
	$H_{\mathrm{O}}$ and $H_{\mathrm{A}}$ are the Hilbert spaces for the cavity and the ancilla mode respectively.
	Thus this operation will entangle the ancilla state $ \left| n \right\rangle $ with the $ n^{\mathrm{th}} $ error subspace.
	
	In the measurement-based QEC, a measurement of the ancilla states is carried out.
	The joint system will be projected onto either the no-error subspace or the subspace with a specific error. Then based on the measurement result, different unitary $U_n $ can be applied to recover the original state.% \red{The recovery operation can be described by:
		%\begin{equation}
		%	{\red{ U_{\mathrm{feed}}}} = \sum\limits_k {U_n } \otimes \left| n \right\rangle \left\langle n \right|.
		%\end{equation}
		%With the application of this operation, the logical states in the error subspaces will be converted back into the code space, and the cavity will get disentangled with the ancilla system.
		%Depending on the errors in the logical state, the ancilla will be left in a state with the error entropy.}
	
	However, it is also possible to finish the error correction without performing the measurement.
	This is realized by using a recovery operation:
	\begin{equation}
		{\red{ U_{\mathrm{recovery}}}} = \sum\limits_n {U_n } \otimes \left| n \right\rangle \left\langle n \right|.
	\end{equation}
	With the application of this operation, the logical states in the error subspaces will be converted back into the code space, and the cavity will get disentangled with the ancilla system.
	Depending on the errors in the logical state, the ancilla will be left in a state with the error entropy.
	%Here lies the difference between QEC implementations between a measurement-based one and an autonomous one.
	The error mapping and the recovery operation can actually be combined into a single one $ {{ U}_{{\mathrm{AQEC}}}} = {{ U}_{{\mathrm{recovery}}}}{{ U}_{{\mathrm{map}}}} $.
	In this way, the requirement of fast real-time electronics at room temperature can be released. Note that to start the next round of QEC iteration, the excitations in the ancilla need to be evacuated.
	
	In the specific example of the lowest binomial code, the logical basis states are defined as
	\begin{equation}
		\left| {{0_L}} \right\rangle  = \frac{{\left| 0 \right\rangle  + \left| 4 \right\rangle }}{{\sqrt 2 }},\ \left| {{1_L}} \right\rangle  = \left| 2 \right\rangle.
	\end{equation}
	The corresponding error basis states are as follows
	\begin{equation}
		\left| {{0_E}} \right\rangle  = \left| 3 \right\rangle,\ \left| {{1_E}} \right\rangle  = \left| 1 \right\rangle.
	\end{equation}
	%So the logical state in the code space could be written as $ \left| {{\psi _{\mathrm{L}}}} \right\rangle  = \alpha \left| {{0_L}} \right\rangle  + \beta \left| {{1_L}} \right\rangle $.
	When an error occurs, the logical state becomes $ \left| {{\psi _E}} \right\rangle  = \alpha \left| {{0_E}} \right\rangle  + \beta \left| {{1_E}} \right\rangle $.
	Thus for this code, the $ {{U}_{{\mathrm{AQEC}}}} $, used for the AQEC should satisfy the following two equations:
	\begin{equation}
		U_{\mathrm{AQEC}}\left| {{\psi _{\mathrm{E}}}} \right\rangle |g\rangle = \left| {{\psi _{\mathrm{L}}}} \right\rangle |e\rangle,
	\end{equation}
	\begin{equation}
		U_{\mathrm{AQEC}}\left| {{\psi _{\mathrm{L}}}} \right\rangle |g\rangle = \left| {{\psi _{\mathrm{L}}}} \right\rangle |g\rangle.
	\end{equation}
	In the first equation, the logical state is recovered from the error subspace and the ancilla is flipped into the first excited state.
	In the second equation, the logical state stays unchanged and the ancilla also remains in its ground state.
	Different from the measurement-based approach, the recovery operation is implemented in a coherent manner.
	When finished, the cavity state becomes disentangled with the ancilla qubit, and the error entropy is transferred onto the ancilla qubit.
	
	In addition to the single-photon loss error, there is also a no-jump evolution $ {e^{ - (\kappa /2){{\hat a}^\dag }\hat at}} $, which will cause the logical states to deform.
	Here $ \kappa $ is the rate for the natural damping and $ t $ is the time duration before the next iteration of the AQEC.
	This error leads to a decreased population of the higher Fock states even if there is no photon-jump event.
	For a small $t$, the no-jump evolution can be approximately corrected by a unitary.
	Hence, we use the following two conditions to restrict our AQEC pulse optimization:
	\begin{equation}
		\begin{aligned}
			{U_{{\mathrm{AQEC}}}}&{U_{{\mathrm{Kerr}}}}\left( {\alpha |3\rangle  + \beta |1\rangle } \right)|g\rangle \\ 
			&= {U_{{\mathrm{Disp}}}^{\dagger}}\left( {\alpha \frac{{|0\rangle  + {e^{{\phi _4}}}|4\rangle }}{{\sqrt 2 }} + \beta {e^{{\phi _2}}}|2\rangle } \right)|e\rangle,\\
			%	\end{aligned}
		%\end{equation}
		%\begin{equation}
		%	\begin{aligned}
			{U_{{\mathrm{AQEC}}}}&{U_{{\mathrm{Kerr}}}}\left( {\alpha \frac{{|0\rangle  + {e^{ - 2\kappa t}}|4\rangle }}{{\sqrt {1 + {e^{ - 4\kappa t}}} }} + \beta |2\rangle } \right)|g\rangle \\ &= \left( {\alpha \frac{{|0\rangle  + {e^{{\phi _4}}}|4\rangle }}{{\sqrt 2 }} + \beta {e^{{\phi _2}}}|2\rangle } \right)|g\rangle.
		\end{aligned}
	\end{equation}
	As before, the first condition is to correct the logical state in the error space and drive the ancilla to an orthogonal state.
	For the second condition, only the no-jump evolution is dealt with.
	Here $ \phi_2 $ and $ \phi_4 $ are the measurement-induced phases for $ \left| 2 \right\rangle $ and $ \left| 4 \right\rangle $, respectively.
	These quantities can be calibrated with an independent experiment.
	$ {U_{{\mathrm{Kerr}}}} $ is the deterministic evolution with the PASS drive between the AQEC operations.
	$ {U_{{\mathrm{Disp}}}^{\dagger}} $ is to cancel the phase accumulation caused by the dispersive frequency shift conditioned on the ancilla qubit being flipped to the excited state, which happens during the measurement and the subsequent waiting process. All the coefficient values for $ \alpha $ and $ \beta $, which are used to optimize the pulse, are shown in \red{Table~\ref{tab:AQEC_GRAPE_coeff}}.
	
	The AQEC pulse is applied on the ancilla qubit and the logical qubit mode simultaneously. The pulse shape is obtained by using gradient ascent pulse engineering (GRAPE), a numerical optimization algorithm~\cite{Khaneja2005JMR, Fouquieres2011JMR}.
	The pulse shape is digitized and parameterized with steps of one nanosecond and the amplitudes at each point are free to be optimized with the experimentally measured parameters. \red{The pulse length is optimized to make the gate fidelity of the AQEC pulse as high as possible when taking into account the decoherence of both the qubit and the cavity.} %chosen to be as short as possible to mitigate the detrimental effect of the decoherence of both the qubit and the cavity.}

\begin{table}
	\centering
	\caption{The coefficients that are used to perform the AQEC GRAPE optimization. The overcomplete set ensures removing the bias for different states.}
	\label{tab:AQEC_GRAPE_coeff}
	\begin{tabular}{p{1cm}<{\centering}|p{1cm}<{\centering}|p{1cm}<{\centering}|p{1cm}<{\centering}|p{1cm}<{\centering}|p{1cm}<{\centering}|p{1cm}<{\centering}} 
		\hline
		Index    & 1   & 2   & 3    & 4    & 5    & 6    \\ 
		\hline
		$\alpha$ & $1$ & $1$ & $1$  & $1$  & $1$  & $0$  \\
		$\beta$  & $0$ & $1$ & $-1$ & $-i$ & $+i$ & $1$  \\
		\hline
	\end{tabular}
\end{table}

After the AQEC operation, the ancilla is in a mixed state, which is caused by the transfer of the error entropy from the logical state to the ancilla.
In order to start the next round of AQEC, the ancilla needs to be reset into its ground state to dump the error entropy.
Typically, there are two general ways to realize it.
The first is the feedback approach, which is based on measurement and feedback operations.
The state of the ancilla qubit is first measured, and then an adaptive qubit drive is applied to flip the qubit if an excited state is measured. Even using this method, the requirement for the control electronics has been greatly released because the reset operation only needs to be completed before the next AQEC operation.
The time window is about several tens of microseconds, and is much less stringent compared with the several hundreds of nanoseconds for the required real-time feedback.
The second method is based on the engineered dissipation of the ancilla qubit.
By applying drives on the ancilla qubit, its coupling with the low-Q readout cavity could be enhanced.
Equivalently the damping time of the ancilla can be shortened by one to two orders of magnitude.
Thus the excitations on the ancilla could be evacuated within several hundreds of nanoseconds~\cite{Magnard2018PRL}.

%The above-discussed constraints are used to estimate the final fidelity of the AQEC pulse .The fidelity is used as the cost function; the stop criteria are set to be within $0.5\% $.

\subsection{Single-side AQEC}

%In the main text, we have shown the continuous AQEC of a Bell state between two logical qubits.

\begin{figure}
	\centering
	\includegraphics[scale=1]{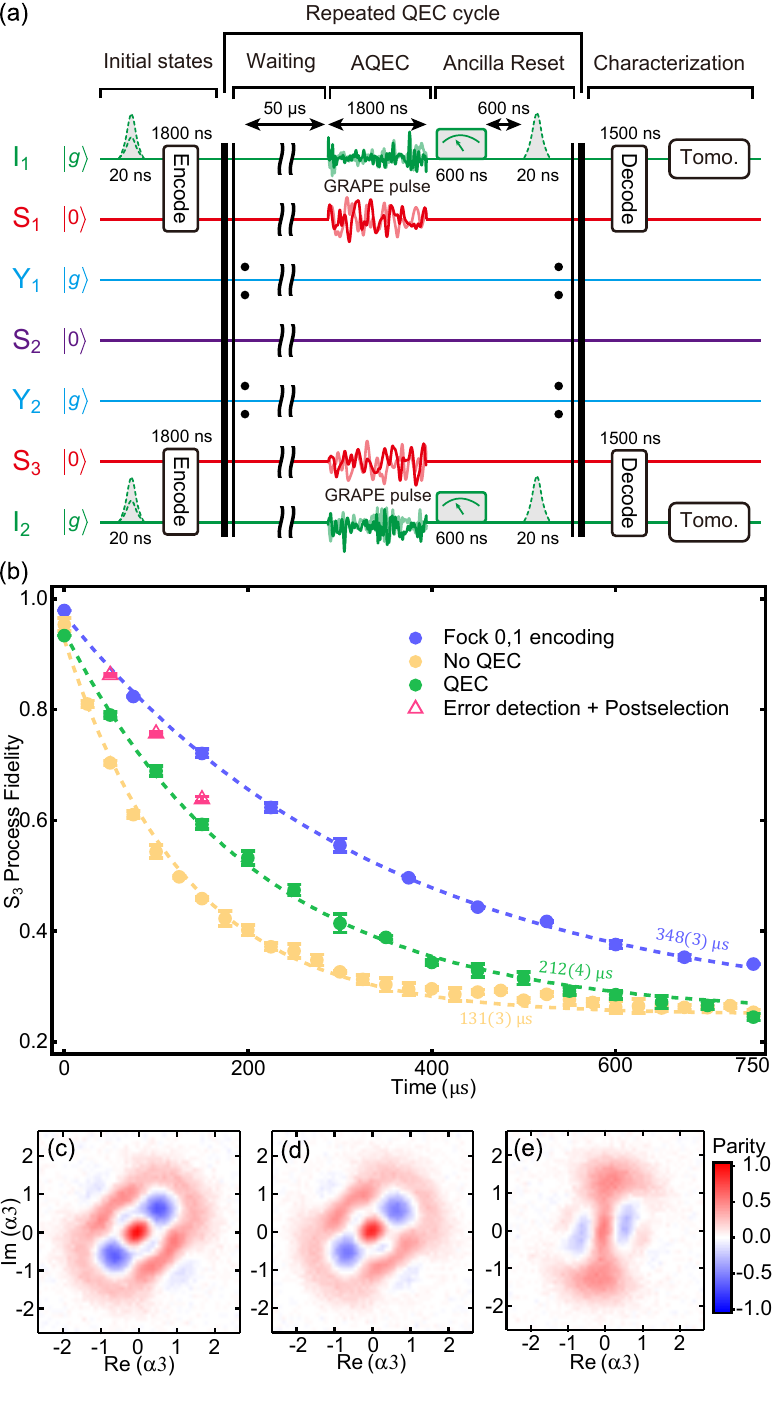}
	\caption{\textbf{The performance of the repetitive AQEC protection on a single binomial code.} (a) The experimental sequence. The repetitive AQEC experiments are performed on both $S_1$ (see the main text) and $S_3$. Here the two experiments are carried out separately, leaving the other side (both the control qubit and the cavity) in the ground states. (b) The performance of the repetitive AQEC protection on $S_3$. The yellow curve shows the binomial code suffering photon loss errors without the protection of the AQEC process. The green curve represents the binomial code protected by repetitive AQEC processes with a time interval of 50~$\mu$s. The process fidelity of the AQEC-protected binomial codes has a significant increase compared to the code without AQEC protection. For comparison, the blue curve represents the process fidelity for Fock state $\{\ket{0}$, $\ket{1}\}$ encoding without any protection. The magenta hollow triangles represent the purified binomial code with post-selection of the no-error case in the experiment under repetitive error syndrome detection but without error correction operations. (c), (d), and (e) show the Wigner function distributions of the binomial state $(\ket{0}+\ket{4})/2-i\ket{2}/\sqrt{2}$ after encoding, at 50~$\mu$s with QEC, and at 50~$\mu$s without QEC, respectively.}
	\label{fig:SingleSide_AQEC}
\end{figure}

%\begin{figure*}
%	\centering
%	\includegraphics[scale=1]{Q5QECdata.pdf}
%	\caption{\textbf{\textcolor{red}{combine with Fig.9}The Process fidelity of different encoding or different quantum error processing of the cavity S3.} The yellow dot line shows the binomial code suffering the photon loss errors without the protection of the AQEC process. The green dot line represents the binomial code protected by repetitive AQEC processes with 50$\mu$s time interval. The process fidelity of the AQEC-protected binomial codes has a huge increase than the code without the AQEC protection. With the experimental data post-processing, the error spaces are dropped from the data of the deterministic AQEC process (green dot line), which makes the experimental results have better performance and is shown as the purple hollow circle. The blue dot line represents the process fidelity changed with the time of Fock 0 and 1 encoding without any protection. The fuchsin hollow triangle represent the binomial code under the repetitive error syndrome detection without error correction operations in the process. The insets at the top, right, and bottom left represent the Wigner tomography of the binomial state ($\ket{0}/2+\ket{2}/\sqrt{2}+\ket{4}/2$), after the encoding, at 50$\mu$s with QEC, and at 50$\mu$s without QEC, respectively.}
%	\label{fig:Q5QECdata}
%\end{figure*}

Here, we provide more detailed experimental data related to the individual AQEC performances of the two logical qubits. The single-side experiments are performed as follows, as shown in Fig.~\ref{fig:SingleSide_AQEC}(a).
$I_1$, $S_1$ and $I_2$, $S_3$ are treated as two independent pairs of basic logical qubit units. Process tomography is utilized to fully characterize the performance of the AQEC protocol. First, different initial states are prepared on the control qubits $I_1$ and $I_2$.
Then these states are encoded onto the binomial bases of $S_1$ and $S_3$, respectively.
Next repetitive AQEC cycles are performed, where the correction pulses are the same as those used in the experiment for the logical Bell state.
Finally, after various rounds of AQEC, the logical states are decoded back onto the control qubits to perform state tomography, which is used to construct the process matrix $ \chi $.
%At this point, the only difference of this protocol, compared with the one used in the main text, is that the Bell state preparation and distribution are replaced with two independent state preparations. 

The results for $S_1$ are shown in the main text. The results for $S_3$ with and without AQEC, as well as the case for Fock state $\{\ket{0},\ket{1}\}$ encoding, are shown in Figs.~\ref{fig:SingleSide_AQEC}(b).
The process fidelity is shown versus the time elapsed, and as expected, both of these curves show exponential decay. For $S_3$, the corrected binomial code has a lifetime of about 1.6 times that of the uncorrected one, demonstrating the protection of logical qubit by AQEC. Figures~\ref{fig:SingleSide_AQEC}(c)-(e) show the Wigner function distributions of the binomial state $(\ket{0}+\ket{4})/2-i\ket{2}/\sqrt{2}$ after encoding, at 50~$\mu$s with QEC, and at 50~$\mu$s without QEC, respectively.

%\subsection{(?) Comparison of the AQEC and conventional measurement-based QEC: fidelity, experimental resources?}

%\section{Experimental Sequence: for the repetitive AQEC}

\begin{figure*}
	\centering
	\includegraphics[scale=1]{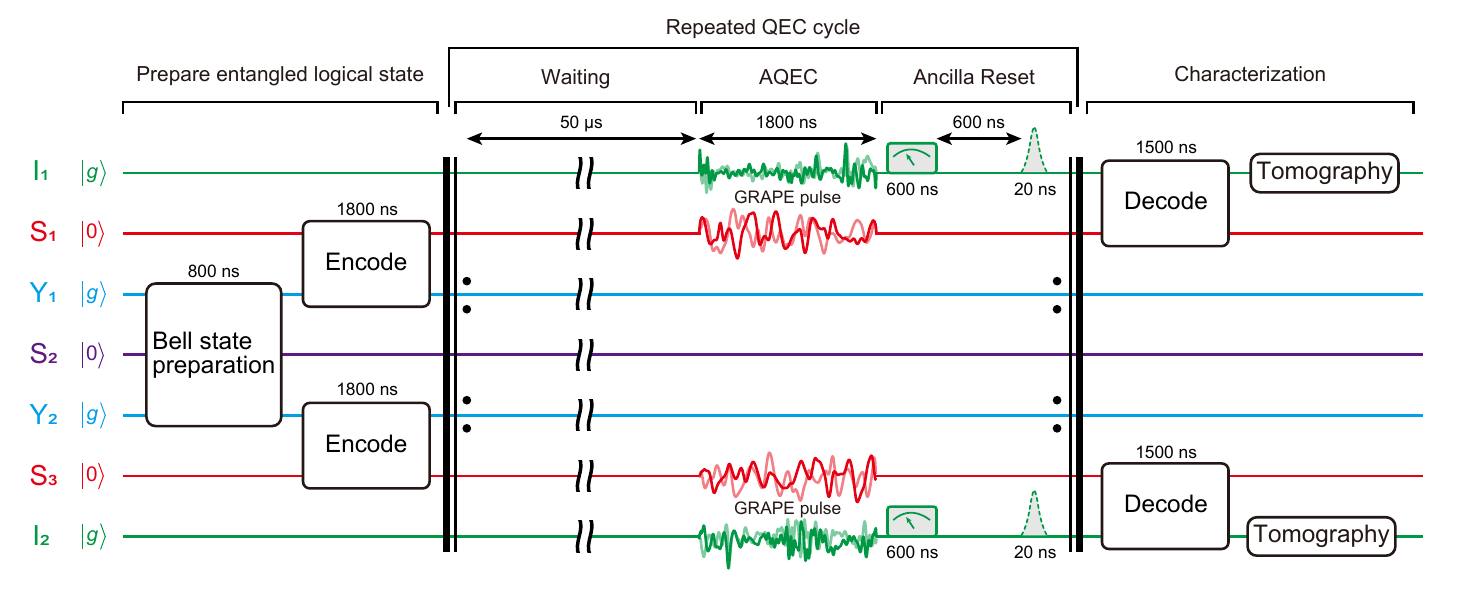}
	\caption{\textbf{Experimental protocol for the repetitive AQEC on the logical Bell state.} %(a) \textcolor{red}{(Put the caption of (a) to fig1?)}The schematic top view of the three-dimensional circuit QED device. $S_1$ and $S_2$, defined as coaxial $\lambda/4$ cavities (red), are used as the two binomial logical qubits. $I_1$ and $I_2$, defined as I-shaped transmon qubits (green), are used to implement the AQEC operation and joint state tomography of the above two logical qubits. In the present work, the two Y-shaped transmon qubits (blue) and the central coaxial $\lambda/4$ cavity $S_3$ (purple), between $S_1$ and $S_2$, are used to provide the necessary effective coupling to generate and distribute Bell states into $S_1$ and $S_2$, as well as enough isolation after the distribution. All the four transmon qubits are dispersively coupled to the adjacent coaxial cavities and a Purcell-filtered $\lambda/2$ stripline readout resonator (black). (b) 
		The protocol for demonstrating the entanglement protection consists of three steps: (1) entanglement generation between $Y_1$ and $Y_2$ assisted by the bus mode $S_2$, and the transfer of the entangled state into $S_1$ and $S_3$; (2) repetitive AQEC operations with a waiting time of 50~$\mu$s and a qubit reset operation; (3) joint two-qubit state tomography after decoding the entangled state back onto the control qubits $I_1$ and $I_2$.}
	\label{fig:AQEC_protocol}
\end{figure*}

\begin{figure*}
	\centering
	\includegraphics[scale=1]{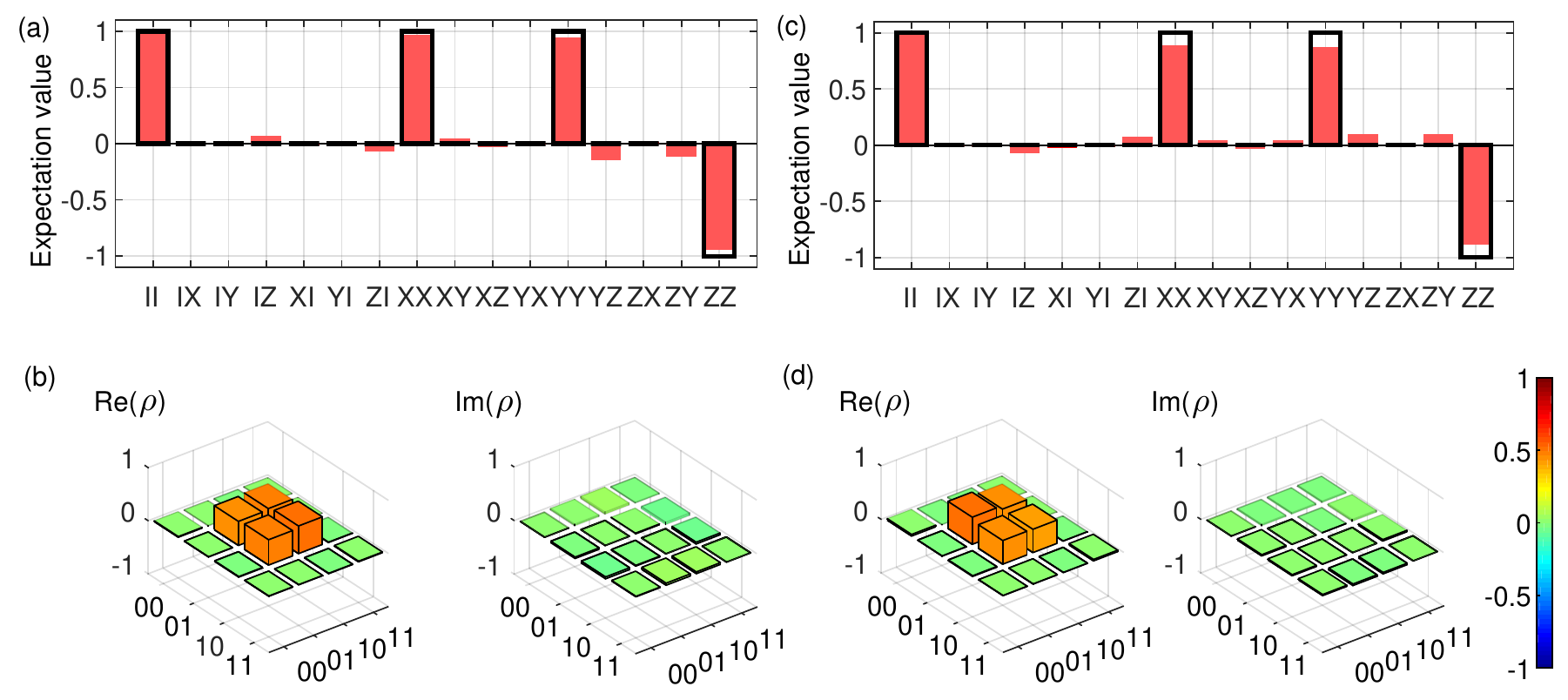}
	\caption{\textbf{Tomography of the Bell state.} (a) Auxiliary qubits $Y_1$ and $Y_2$ are prepared in the Bell state $(|01\rangle +|10\rangle)/\sqrt2 $ and characterized with a joint qubit measurement. (b) Reconstructed density matrix of the Bell state between the auxiliary qubits $Y_1$ and $Y_2$ shows a state fidelity of 0.965. (c) The $Y_1$-$Y_2$ entangled state is encoded into the logical qubits $(|0_{\mathrm{L}}1_{\mathrm{L}}\rangle +|1_{\mathrm{L}}0_{\mathrm{L}}\rangle)/\sqrt2 $, which is decoded back onto the control qubits $I_1$ and $I_2$, followed by a joint qubit measurement. (d) Reconstructed density matrix of the Bell state between the control qubits $I_1$ and $I_2$, \red{inferring the fidelity of the logical Bell state is 0.912 (with the encode, decode, and measurement errors included)}.}
	\label{fig:SM_Q4Q6EN}
\end{figure*}

\subsection{Repetitive AQEC on the logical Bell state}

The cross-talk between $S_1$ and $S_3$ can be safely ignored because cavity $S_2$ provides enough isolation between them.
Then the two individual AQEC experiments can be performed at the same time. In the main text, we have shown the continuous AQEC of a Bell state of two logical qubits. The experimental protocol is shown in Fig.~\ref{fig:AQEC_protocol}. The protocol for demonstrating entanglement protection consists of three steps:

(1) Preparation of entangled logical state. An entanglement between the auxiliary qubits $Y_1$ and $Y_2$ is first generated with the assistance of the bus mode $S_2$. The fidelity of this Bell state is 0.965 based on the measured reconstructed density matrix, as shown in Fig.~\ref{fig:SM_Q4Q6EN}(a) and \ref{fig:SM_Q4Q6EN}(b). Then the entangled state between $Y_1$ and $Y_2$ is encoded onto the two logical qubits $S_1$ and $S_3$. The fidelity of the prepared logical Bell state is measured to be 0.912, as shown in Figs.~\ref{fig:SM_Q4Q6EN}(c) and \ref{fig:SM_Q4Q6EN}(d).

(2) Repetitive AQEC operations. A waiting time of 50~$\mu$s is chosen between consecutive AQEC operations to balance the AQEC operation errors and the uncorrectable errors of the code (see Sec.~\ref{sec:ErrorAnalysis} for details). After each AQEC operation, a qubit reset is applied to eliminate the error entropy for the next round of AQEC. This whole process is repeated variable times.

(3) Characterization. Finally the entangled logical state is decoded back onto the control qubits $I_1$ and $I_2$, followed by joint two-qubit state tomography to extract the final Bell state fidelity.

\section{Numerical simulation of the AQEC performance}

\subsection{Numerical results}
To understand the limiting factors for the performance of AQEC in our system, we implement numerical simulation in QuTiP with decoherence and system Hamiltonian parameters calibrated in the experiment~\cite{Qutip2013}. In our simulation, the dimension of the cavity is truncated to 30, and the control qubit is considered as an ideal two-level system. Even so, simulation of the whole system still requires tremendous time and calculation resources. Since the cross Kerr between $S_1$ and $S_3$ is negligible, we assume that the AQEC processes of the two logical qubits are independent. Therefore, for efficiency, we first simulate the evolution of each logical qubit separately and obtain the corresponding process matrices. Then, we use the tensor product of matrices from the two sides as the two-logical-qubit process matrix and act them on the entangled state to get the final results. To verify our method, we apply the process matrices obtained from the experiment on the entangled state, and the evolution curve we get from this method agrees well with the experiment.

\begin{figure}
	\centering
	\includegraphics[scale=1]{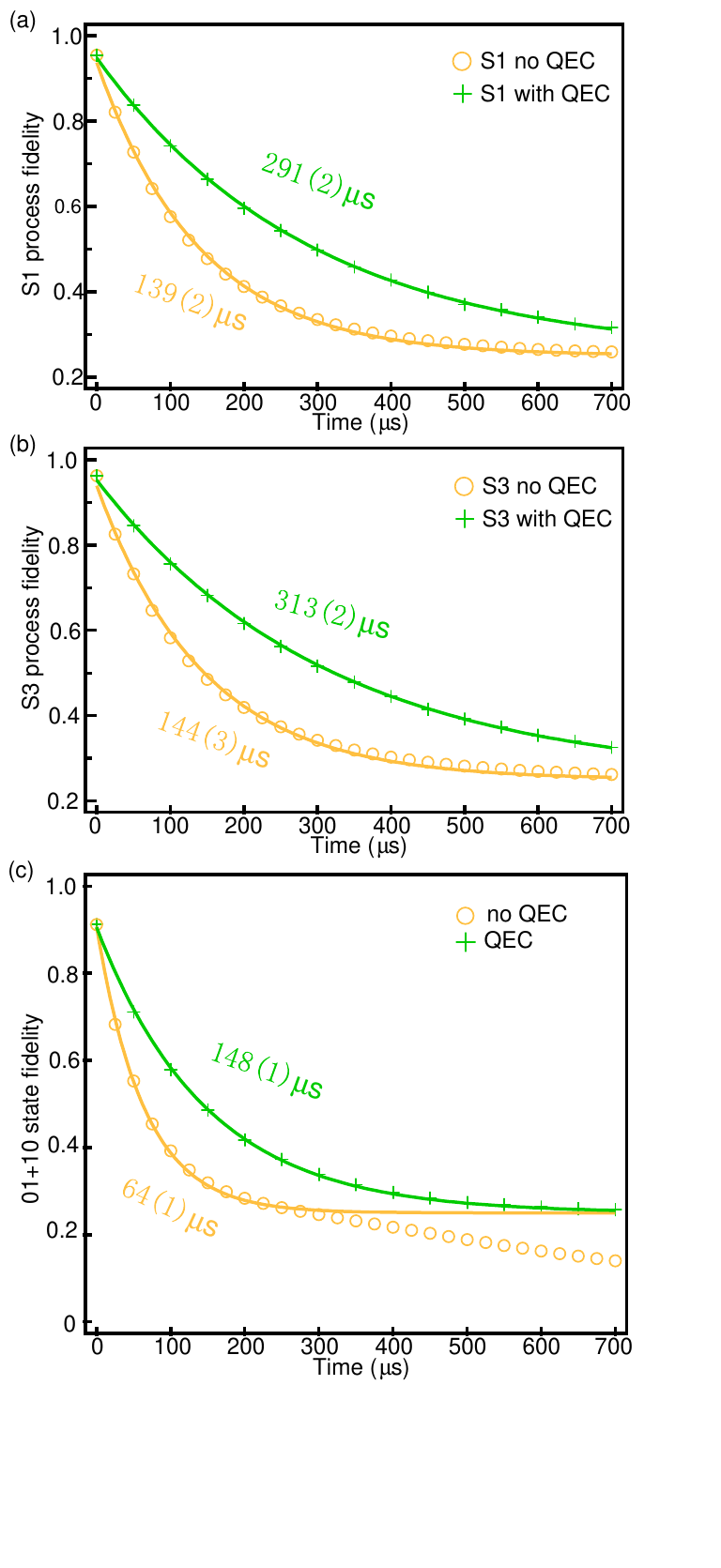}
	\caption{\textbf{Numerical simulation results for the two logical qubits and the Bell state with and without AQEC.} All curves are fitted using $F(t)=0.25+Ae^{-t/T} $ to obtain the decay time of the process fidelity . (a) and (b) show the process fidelity of the two logical qubits as a function of time separately. (c) is obtained by implementing the process matrices from (a) and (b) to the Bell state.}
	\label{fig:Simulation}
\end{figure}

Using the approach described above we get the simulation results of each logical qubit and the entangled state with and without AQEC, as shown in Fig.~\ref{fig:Simulation}. It should be noted that in addition to the decoherence of qubits and cavities, we also add measurement errors to the simulation by mixing the code space and error space according to the measured readout fidelities in Table~\ref{tab:SM_readout_Fidelity}. Comparing the simulation results with those of the experiment, we find that the simulation results of $S_1$ are in good agreement with the experiment, while for $S_3$ there is an obvious deviation. We observe in the experiment that the parameters of qubit $I_2$ fluctuate greatly with time, which affects the performance of the AQEC gate. We attribute the fluctuation to unstable two-level systems.

\subsection{Error analysis}
\label{sec:ErrorAnalysis}
To further understand the error budget of our system, we carry out more analyses and simulations. First, we divide the errors into two categories. The first kind of error occurs during the repetitive waiting and AQEC process, including uncorrectable errors, AQEC gate errors, and measurement-induced errors. This kind of error will affect the efficiency of QEC and determine the lifetime of the logical-qubit entanglement directly. The other kind of error, including non-ideal initial states and encode-decode errors, only contributes an offset to the total fidelity. In this section, we will only focus on the first kind of error.

Based on the process matrices in 
the simulation and experiment, \red{we find that the process after a period of waiting time followed by an AQEC gate can be approximated as a depolarization channel}, which can be described by
\begin{equation}
	\label{eq:depolarizing}
	\varepsilon(\rho)=(1-p)\rho+pI/2,
\end{equation}
where $I$ is the identity matrix and $\rho$ is the density matrix. By implementing such a channel repetitively, the decay time of the process can be expressed as
\begin{equation}
	T=-\tau/\mathrm{ln}(1-p),
\end{equation}
where $\tau$ is the length of the waiting time between AQECs. For simplification, we also assume that different errors are independent, that is 
\begin{equation}
	1-p=\prod_{i=1}^3(1-p_i),
\end{equation}
where $i$ stands for different errors, including uncorrectable errors, AQEC gate errors, and measurement-induced errors. In the following, we will analyze these three types of errors separately, and calculate their contributions to the total fidelity $1-p$. 
\begin{figure}
	\centering
	\includegraphics[scale=1]{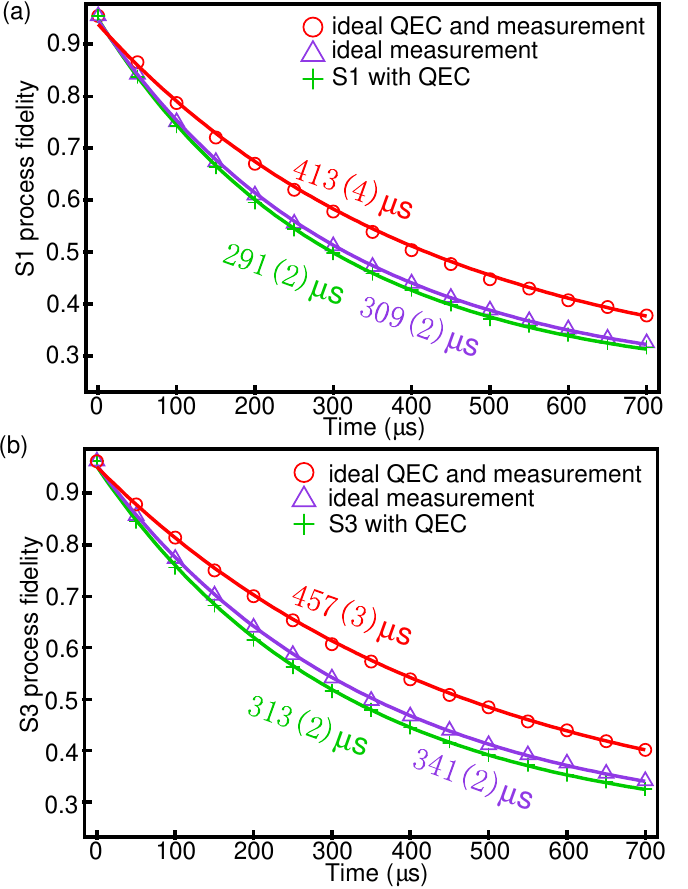}
	\caption{\textbf{The AQEC simulation results for different errors.} All curves are fitted using $F(t)=0.25+Ae^{-t/T} $ to obtain their decay times. The green curves are obtained from simulations with all types of errors. The purple curves are from simulations without measurement errors. The red curves are from simulations with neither measurement errors nor AQEC gate errors.}
	\label{fig:erroranalysis}
\end{figure}

\subsubsection{AQEC gate error}
As described in section~\ref{sec:Autonomous quantum error correction}, the AQEC gates are realized with microwave pulses optimized by the GRAPE algorithm. The numerical optimizations have converged well, with an infidelity smaller than $0.5\%$. However, the performance in the experiment is also limited by the decoherence of the control qubit and the cavity. Since the GRAPE pulse duration is much shorter than the coherence times of our system, the dependence of the fidelity on the coherence is expected to be linear. By changing the coherence parameters in the simulation and performing linear fitting, we find that the fidelities of the AQEC gates for $S_1$ and $S_3$ satisfy
\begin{equation}
	F_{S_1}=0.9977*(1-0.848/T_1)(1-0.745/T_{\phi})(1-3.73/T_c)
\end{equation}
and
\begin{equation}
	F_{S_3}=0.9979*(1-0.735/T_1)(1-0.765/T_{\phi})(1-3.71/T_c),
\end{equation}
where $T_1$ and $T_{\phi}$ are the corresponding control qubit lifetime and dephasing time, respectively, and $T_c$ is the lifetime of the cavity. The coefficients 0.9977 and 0.9979 are the final fidelities for the numerical optimization of the AQEC pulses. It is noted that the fidelities here represent the average fidelities of the optimization conditions, and we find that they are approximately equal to $(1-p)_\mathrm{AQEC}$. Applying the parameters that we used to get the curves in Fig.~\ref{fig:Simulation}, we can get the values of $(1-p)_\mathrm{AQEC}$, as shown in Table~\ref{tab:Errorbudget}.

\subsubsection{Measurement errors}
Because we use measurement and feedback to reset the control qubit after AQEC operations, measurement errors erroneously indicating the control qubit state will cause the control qubit to end up in the excited state, and finally propagate errors to the logical state in the cavity. As mentioned before, we add measurement errors to our simulation by mixing the code space and error space according to the measurement fidelity in Table~\ref{tab:SM_readout_Fidelity}. If we eliminate this kind of error in the simulation, the \red{process fidelity decay times} of $S_1$ and $S_3$ change from 291~$\mu$s and 313~$\mu$s to 309~$\mu$s and 341~$\mu$s, respectively, as shown by the green and purple curves in Fig.~\ref{fig:erroranalysis}. After calculating $(1-p)$ with and without measurement errors, we can make a simple division and get the contributions of measurement errors, shown as $(1-p)_\mathrm{Measure}$ in Table~\ref{tab:Errorbudget}.

\subsubsection{Uncorrectable errors}
The lowest-order binomial code can only correct single-photon-loss error. Although single-photon loss is the dominant error, other errors such as dephasing and higher-order photon loss also limit the performance of the QEC process. To extract the influence of these uncorrectable errors, we run the simulation with the ideal AQEC gate and exclude the measurement errors. As described above, the AQEC GRAPE pulses without decoherence are very close to ideal AQEC gates. Therefore, for convenience, we exclude decoherence when implementing the AQEC pulses instead of using ideal gates. The results show that the process fidelity decay times of $S_1$ and $S_3$ change to 413~$\mu$s and 457~$\mu$s respectively, as shown in Fig.~\ref{fig:erroranalysis}. In order to eliminate the intrinsic errors of the repetitive AQEC pulses, we divide $(1-p)$ obtained from the relation $T=-\tau/\mathrm{ln}(1-p)$ by 0.9977 and 0.9979. The final $(1-p)_\mathrm{Uncorrectable}$ values are shown in Table~\ref{tab:Errorbudget}.

\begin{table}
	\centering
	\caption{The simulated error budget. $(1-p)$ is defined by Eq.~\ref{eq:depolarizing}.}
	\label{tab:Errorbudget}
	\begin{tabular}{c|cc} 
		\hline
		Error type         & $S_1$ & $S_3$   \\ 
		\hline
		$(1-p)_\mathrm{AQEC}$   & 95.9\%           & 96.6\%                          \\
		$(1-p)_\mathrm{Uncorrectable}$         & 88.8\%           & 89.8\%                          \\
		$(1-p)_\mathrm{Measure}$        & 99\%           & 98.7\%                           		\\
		$(1-p)_\mathrm{total}$         & 84.2\%           & 85.2\%                              \\
		\hline
	\end{tabular}
\end{table}

% \section{Numerical simulation of the AQEC performance?}

% \subsection{Qutip results?}

% \subsection{Error analysis or Error Budget(gate infidelity)?}

\section{Concurrence, negativity, and CHSH inequality of the entangled states}
\textbf{Concurrence}. Given a reconstructed density matrix, metrics indicating the quality of the corresponding experimental state can be calculated. In addition to the state fidelity, a more direct metric to quantify the degree of entanglement between two qubits is needed. Usually, the quantum monotone is used in practice~\cite{Horodecki2009RMP}. It is a monotonic function, with its value being 0 for a separable state and 1 for a Bell state, e.g. the maximally entangled state.
The quantum monotone remains unchanged under any given local unitary operation, and will not increase by the combination of any local operation and classical communication channels.
The general computational process is to first reconstruct the system density matrix $\rho$ with maximum-likelihood estimation (MLE), and then calculate the quantum monotone from the eigenvalues of the related matrices.

A specific example is concurrence, which is widely used in quantum information experiments.
Once the density matrix $\rho$ has been reconstructed, the \red{``spin-flipped state"} first needs to be calculated as $\tilde{\rho}=\left(\sigma_{y} \otimes \sigma_{y}\right) \rho^{*}\left(\sigma_{y} \otimes \sigma_{y}\right)$, where $\sigma_{y}$ is one of the Pauli matrices and the star symbol represents the complex conjugate.
Then the concurrence matrix is obtained using the relation $R=\sqrt{\sqrt{\rho} \tilde{\rho} \sqrt{\rho}}$.
Finally, the concurrence can be extracted by the equation $\mathcal{C}(\rho) \equiv \max \left\{0, \lambda_{1}-\lambda_{2}-\lambda_{3}-\lambda_{4}\right\}$, where $\lambda_{1},\lambda_{2},\lambda_{3},\lambda_{4}$ are the eigenvalues of the concurrence matrix $R$ in decreasing order.
The value for concurrence is invariant under different choices of the basis of $\rho$. Any state with $\mathcal{C}>0$ indicates that entanglement exists. 

\textbf{Negativity}. Negativity is another example of entanglement monotones, which is easy to compute and proper to be a gauge of the degree of entanglement~\cite{Vidal2002PRA}.
It is defined as
\begin{equation}
	\mathcal{N}(\rho ) = \frac{{{\mathrm{tr}}\sqrt {{{\left( {{\rho ^{{{\mathrm{T}}_{\mathrm{A}}}}}} \right)}^\dagger }{\rho ^{{{\mathrm{T}}_{\mathrm{A}}}}}}  - 1}}{2},
\end{equation}
where $\rho  = {\rho _{{\mathrm{A}} \otimes {\mathrm{B}}}}$ is the density matrix of the joint system of subsystem A and subsystem B, and ${{\rho ^{{{\mathrm{T}}_{\mathrm{A}}}}}}$ denotes taking the partial transpose of $\rho$ with respect to subsystem A.

\textbf{Bell signal}. Traditionally, the Bell test is designed to test the correctness of quantum theory~\cite{Bell1964PPF}.
In this work, we use it as a metric of entanglement to characterize the degree of entanglement of the logical Bell state and to show the effectiveness of the entanglement protection protocol.
A Bell test often involves demonstrating the violation of the Bell inequality and seeking the maximal value of the correlation measurement.
CHSH inequality is one of the most widely used Bell tests, because it has a well-defined classical bound, $ \langle {\mathcal B}\rangle = 2 $, where $ {\mathcal B} $ is the Bell operator~\cite{Clauser1969PRL}.
Quantum theory predicts that an upper bound for this observable is $ \langle {\mathcal B}\rangle = 2\sqrt{2} $, which is achievable by maximally entangled states.
With $ \langle {\mathcal B}\rangle > 2 $, the prepared state must be entangled and inseparable, and the value indicates the degree of entanglement. However, with any measurement result of $ \langle {\mathcal B}\rangle < 2 $, the prepared state cannot be determined to be entangled or separable.

The Bell operator that we use is defined as
\begin{equation}
	\mathcal{B} = QS + RS + RT - QT,
\end{equation}
where $ \{Q=-Y,\ R=X\} $ are the two different measurements after decoding the $S_1$ state onto $I_1$, and $ \{S=(Y-X)/\sqrt{2},\ T=(-Y-X)/\sqrt{2}\} $ are those after decoding the $S_3$ state onto $I_2$.
$ \{X,Y,Z\} $ are the Pauli operators.
For any single measurement, 1 or -1 is assigned to $ Q,R,S,T $ when the corresponding qubit is measured to be at $ \left| g \right\rangle $ or $ \left| e \right\rangle $.
In our experiment, to get the maximal violation of the Bell test, we keep the measurement axes fixed and execute a single-logical-qubit rotation to the prepared Bell state.
This is implemented by changing the rotation axis of $Y_1$ in the encoding GRAPE pulse. 
% which equals the operation
%\begin{equation}
%	\red{R_L^z(\theta ) \otimes {I_L}}.
%\end{equation}
For the above-chosen measurement and the prepared Bell state, the maximal violation of the CHSH inequality corresponds to the single-logical-qubit rotation of angle $ n\pi $ (as shown in Figs.~4(a) and 4(b) in the main text), with $n$ being integers.

%==========================================
%Bell operator: 

%$ \mathcal{B} = QS + RS + RT - QT $

%$ Q=-Y, R=X $

%$ S=(Y-X)/\sqrt{2}, T=(-Y-X)/\sqrt{2} $

\section{Joint Wigner function}

To reconstruct the density matrix of the bosonic state $\rho$, the Wigner function is first measured.
Wigner tomography is a widely used method to fully characterize the quantum state of a bosonic mode and it is the quasi-probability distribution in the phase space $\mathrm{Re}(\beta)$-$\mathrm{Im}(\beta)$.
It is performed by measuring the photon-number parity operator $\hat P = {e^{i\pi {{\hat a}^\dag }\hat a}}$ of the target mode after different displacement operations in the two-dimensional phase space
\begin{equation}
	\begin{aligned}
		W\left( \beta  \right) &= \frac{2}{\pi }{\mathop{\mathrm Tr}\nolimits} \left[ {\rho {{\hat D}_\beta }\hat P\hat D_\beta ^\dag } \right]\\
		&= \frac{2}{\pi }P(\beta),
	\end{aligned}
\end{equation}
where ${{\hat D}_\beta } = {e^{\beta {{\hat a}^\dagger } - {\beta ^*}\hat a}}$ is the displacement operation on the target mode by an amplitude of $\beta$.
The parity measurement is implemented by a Ramsey-type experiment on the adjacent control qubit with a waiting time of $t=\pi/\chi_{\mathrm{sq}}$, where $\chi_{\mathrm{sq}}$ is the dispersive interaction strength~\cite{Sun2014Nature}.
After measuring the Wigner function, an MLE can be performed to extract the most likely density matrix $\rho$, with which the state fidelity can be calculated.
For individual logical qubit $S_1$ ($S_3$), the state fidelity after encoding is 0.9383 (0.9310), which is limited by the decoherence of both the control qubit and the storage cavity during the encoding process.
After $50~\mu$s of free evolution, the state fidelity for $S_1$ ($S_3$) becomes 0.4567 (0.4160).
While, with the AQEC operation after the $50~\mu$s of free evolution, the state fidelity is greatly improved to 0.8564 (0.8057), which demonstrates the effectiveness of the QEC protection of individual logical qubits.

Nevertheless, the single mode Wigner tomography cannot be directly applied to characterize the quantum state of entangled logical qubits, because it does not reveal the correlation between the two modes.
Here we use the joint Wigner tomography~\cite{Wang2016Science} to examine the performance of the entanglement protection protocol, as shown in the insets of Fig.~3 in the main text.
It is obtained by measuring the displaced joint photon number parity of the two logical qubits
\begin{equation}
	\begin{aligned}
		{W_J}\left( {{\beta _1},{\beta _2}} \right) &= \frac{4}{{{\pi ^2}}}{\mathop{\mathrm Tr}\nolimits} \left[ {\rho {{\hat D}_{{\beta _1}}}{{\hat D}_{{\beta _2}}}{{\hat P}_J}\hat D_{{\beta _2}}^\dag \hat D_{{\beta _1}}^\dag } \right] \\
		&= \frac{4}{{{\pi ^2}}}{P_J}\left( {{\beta _1},{\beta _2}} \right),
	\end{aligned}
\end{equation}
where ${\hat P_J} = \exp [i\pi (\hat a_1^\dagger {\hat a_1} + \hat a_2^\dagger {\hat a_2})] = {\hat P_1}{\hat P_2}$ is the joint photon number parity operator, ${{\hat D}_{\beta_{1(2)}} }$ is the displacement operation on the 1st (2nd) logical qubit, and $\rho$ is the two-mode state to be characterized. The second equation holds because the operators of bosonic modes in different cavities commute with each other. As can be seen, the joint Wigner tomography is a function in a 4-dimensional phase space $\mathrm{Re}(\beta_1)$-$\mathrm{Im}(\beta_1)$-$\mathrm{Re}(\beta_2)$-$\mathrm{Im}(\beta_2)$.
The measurement of the joint photon number parity is implemented by simultaneously carrying out Ramsey-type experiments on the individual corresponding control qubits.
To visualize the correlation in the entangled states, 2D cuts along certain planes are shown in the main text.

%==========================================

%For cavity S1 or S3, the fidelity of reconstructed density %matrix from Wigner tomography.
%
%after encode: S1: 0.9383,  S3: 0.9310
%
%after 50$\mu s$ without AQEC: S1: 0.4567, S3: 0.4160
%
%after 50$\mu s$  with AQEC: S1: 0.8564, S3: 0.8057

\section{Error detection and purification}

The purification strategy enables better performance for single-cavity process tomography or two-cavity entanglement protection via error detection. The cost is the reduction of the amount of available data by discarding the experimental trajectories with any photon-jump errors.
The procedure is as follows.
First, with a specific interval of time, the parity measurement protocol is applied repetitively on the logical qubits, with the pulse sequence of $(R_{\pi /2}^Y,\pi /{\chi_\mathrm{sq}},R_{ - \pi /2}^Y)$. The measurement result will be recorded, with the control qubit in $\left| g \right\rangle$ indicating that there is no error occurring.
In the end, the bosonic state is decoded onto the corresponding control qubit with the no-jump error being corrected in the decoding pulse. With this protocol, a sequence of qubit measurement results is available. At each time interval, choosing only the data corresponding to an even parity state (with no photon-jump error) gives the post-selected data for purified logical qubits.

%Control: production of eprint (-1) disabled
%

%\bibliographystyle{Zou}
%\bibliography{SM_refs}

\end{document}